\documentclass[10pt,twocolumn,aps,prx,superscriptaddress,longbibliography,eprint]{revtex4-1}

\usepackage{graphicx}
\usepackage{amsmath}
\usepackage{amssymb}
\usepackage{float} 
\usepackage{braket} 
\usepackage{enumitem}
\usepackage[dvipsnames]{xcolor}
\usepackage[colorlinks=true, urlcolor=blue, linkcolor=blue,citecolor=blue, hypertexnames=false, unicode=true]{hyperref}
\usepackage[detect-weight=true,separate-uncertainty=true]{siunitx}
\usepackage{comment}

\usepackage{cleveref}
\crefname{equation}{Eq.}{Eqs.}
\Crefname{equation}{Equation}{Equations}
\crefname{figure}{Fig.}{Figs.}
\Crefname{figure}{Figure}{Figures}
\crefname{section}{Sec.}{Sects.}
\Crefname{section}{Section}{Sections}
\crefname{table}{Table}{Tables}
\crefname{appendix}{Appendix}{Apps.}
\Crefname{appendix}{Appendix}{Apps.}

\newcommand{\ahat}{\hat{a}}
\newcommand{\nm}[2]{(#1\!:\!#2)}

\makeatletter
\newcommand{\dashover}[2][\mathop]{#1{\mathpalette\df@over{{\dashfill}{#2}}}}
\newcommand{\fillover}[2][\mathop]{#1{\mathpalette\df@over{{\solidfill}{#2}}}}
\newcommand{\df@over}[2]{\df@@over#1#2}
\newcommand\df@@over[3]{
 \vbox{
 \offinterlineskip
 \ialign{##\cr
  #2{#1}\cr
  \noalign{\kern1pt}
  $\m@th#1#3$\cr
 }
 }
}
\newcommand{\dashfill}[1]{
 \kern-.5pt
 \xleaders\hbox{\kern.5pt\vrule height.4pt width \dash@width{#1}\kern.5pt}\hfill
 \kern-.5pt
}
\newcommand{\dash@width}[1]{
 \ifx#1\displaystyle
 2pt
 \else
 \ifx#1\textstyle
  1.5pt
 \else
  \ifx#1\scriptstyle
  1.25pt
  \else
  \ifx#1\scriptscriptstyle
   1pt
  \fi
  \fi
 \fi
 \fi
}
\newcommand{\solidfill}[1]{\leaders\hrule\hfill}
\makeatother

\begin{document}

\title{Measurement-Induced Transmon Ionization}

\author{Marie Fr\'ed\'erique Dumas}
\thanks{These authors contributed equally.}
\affiliation{Institut Quantique and D\'epartement de Physique, Universit\'e de Sherbrooke, Sherbrooke J1K 2R1 QC, Canada}
\author{Benjamin Groleau-Par\'e}
\thanks{These authors contributed equally.}
\affiliation{Institut Quantique and D\'epartement de Physique, Universit\'e de Sherbrooke, Sherbrooke J1K 2R1 QC, Canada}
\author{Alexander McDonald}
\affiliation{Institut Quantique and D\'epartement de Physique, Universit\'e de Sherbrooke, Sherbrooke J1K 2R1 QC, Canada}
\author{Manuel H. Mu\~noz-Arias}
\affiliation{Institut Quantique and D\'epartement de Physique, Universit\'e de Sherbrooke, Sherbrooke J1K 2R1 QC, Canada}
\author{Crist\'obal Lled\'o}
\affiliation{Institut Quantique and D\'epartement de Physique, Universit\'e de Sherbrooke, Sherbrooke J1K 2R1 QC, Canada}
\author{Benjamin D'Anjou}
\affiliation{Institut Quantique and D\'epartement de Physique, Universit\'e de Sherbrooke, Sherbrooke J1K 2R1 QC, Canada}
\author{Alexandre Blais}
\affiliation{Institut Quantique and D\'epartement de Physique, Universit\'e de Sherbrooke, Sherbrooke J1K 2R1 QC, Canada}
\affiliation{Canadian Institute for Advanced Research, Toronto, M5G 1M1 Ontario, Canada}

\date{\today}

\begin{abstract}
 Despite the high measurement fidelity that can now be reached, the dispersive qubit readout of circuit quantum electrodynamics is plagued by a loss of its quantum nondemolition character and a decrease in fidelity with increased measurement strength. In this work, we elucidate the nature of this dynamical process, which we refer to as transmon ionization. We develop a comprehensive framework which provides a physical picture of the origin of transmon ionization. This framework consists of three complementary levels of descriptions: a fully quantized transmon-resonator model, a semiclassical model where the resonator is treated as a classical drive on the transmon, and a fully classical model. Crucially, all three approaches preserve the full cosine potential of the transmon and lead to similar predictions. This framework identifies the multiphoton resonances responsible for transmon ionization. It also allows one to efficiently compute numerical estimates of the photon number threshold for ionization, which are in remarkable agreement with recent experimental results. The tools developed within this work are both conceptually and computationally simple, and we expect them to become an integral part of the theoretical underpinning of all circuit QED experiments.
\end{abstract}

\maketitle

\section{Introduction}

Circuit quantum electrodynamics with transmon qubits is a leading platform for quantum computation~\cite{Blais2021Circuit,Koch2007ChargeInsensitive}. A key feature of this architecture is the dispersive readout, where the qubit state is inferred from quadrature measurement of a microwave signal transmitted or reflected from a resonator coupled to the qubit~\cite{Wallraff2005}. In principle, increasing the amplitude of the measurement drive, or equivalently the average resonator photon population, should lead to an increase in the measurement's signal-to-noise ratio (SNR). Moreover, because the qubit is strongly detuned from the resonator and, thus, from the readout drive, the process is expected to be quantum nondemolition (QND)~\cite{Blais2004Cavity}. 

In practice, however, increasing the readout drive amplitude results in a reduction of the measurement fidelity and in measurement-induced transitions spoiling the QND character of the readout. This is a widely observed phenomenon~\cite{Walter2017Rapid,Sank2016Measurement,Khezri2023Measurement,Minev2019,Swiadek2023,Johnson2012}. Although it has been experimentally possible to optimize the dispersive readout to obtain high-fidelity state discrimination~\cite{Jeffrey2014Fast,Walter2017Rapid,Touzard2019,Sunada2022,Lledo2023Cloaking,Chen2023Transmon,Swiadek2023}, single-shot qubit readout now lags in fidelity behind the best one- and two-qubit gates~\cite{Kjaergaard2020Superconducting}. Thus, improving qubit readout is an outstanding challenge for the field. Attempts at understanding the origin of these observations have been made using perturbative expansions in the qubit-resonator coupling, qubit anharmonicity, or drive amplitudes while treating the qubit as a two-level system or a Kerr-nonlinear oscillator~\cite{Boissonneault2008,Boissonneault2009,Petrescu2020,Hanai2021}. While these approaches highlight important mechanisms for qubit decay and excitation in circuit QED, they do not explain the observed non-QNDness of the dispersive readout. This situation is exacerbated by the fact that no clear parameter dependence for the onset of non-QNDness emerges from experimental observations. In a different line of work, the accidental presence of a two-level defect near the qubit transition frequency was shown to explain some experimentally observed features~\cite{Thorbeck2024ReadoutInduced}. However, this does not account for the multilevel nature of the measurement-induced transitions observed in many experiments.

There has recently been a flurry of activity aimed at understanding the origin of the failure of the dispersive readout~\cite{Shillito2022Dynamics,Cohen2023Reminiscence,Khezri2023Measurement,Sank2016Measurement,Xiao2023Diagrammatic,Verney2019,Lescanne2019Escape}. In particular, numerical simulations of the dynamics of the dispersive readout have shown that the measurement drive, although strongly detuned from the qubit in the dispersive regime, can cause measurement-induced transitions to highly excited states of the transmon at specific resonator photon numbers~\cite{Shillito2022Dynamics}. In some cases, this leakage extends to states beyond the transmon's cosine potential well, a phenomenon referred to as transmon ionization. Moreover, \textcite{Cohen2023Reminiscence} showed that the chaotic behavior of the driven nonlinear classical pendulum can be used to gain qualitative understanding of transmon ionization, an observation which suggests that perturbative methods may not be sufficient to understand ionization. In view of the transmon's negative anharmonicity, the importance of the qubit-resonator frequency detuning, i.e., placing the transmon frequency above or below that of the resonator, was highlighted in Refs.~\cite{Khezri2023Measurement,Cohen2023Reminiscence}.

However, it is unclear how these seemingly disparate approaches relate to each other and whether they make the same quantitative or even qualitative predictions. In this work, we present a comprehensive picture of ionization in the transmon qubit. To do so, we divide the analysis into three main levels of approximation, in all cases considering both negative and positive detunings. We first consider a fully quantized transmon-resonator model that enables us to investigate the physical mechanisms leading to ionization. We then argue that, under typical conditions, a simplified semiclassical model where the resonator is treated as a classical field within Floquet theory reproduces the features of the full quantum model and is able to accurately predict ionization. Because it omits the resonator mode, this model is much more numerically efficient without losing predictive power. Finally, we demote the transmon to a fully classical system and use the Bohr-Sommerfeld quantization rule to make predictions about ionization. Remarkably, we show that all three models lead to similar qualitative and quantitative predictions for the onset of ionization in a wide range of qubit-resonator detunings and that they agree with experimental results. We also highlight the importance of gate charge in the onset of ionization of the computational states, even deep in the transmon regime~\cite{Cohen2023Reminiscence,Khezri2023Measurement}. Common to all three approaches is that they consider the full cosine potential of the transmon rather than the common Kerr nonlinearity approximation. 

An important feature emerging from all three methods is that, in general, there exist several critical photon numbers at which ionization can occur in principle. These critical points correspond to resonances that are activated by the strong resonator field
thanks to the Josephson junction's nonlinearity. We show that not all these resonances play an equally important role and that some can be safely ignored in the appropriate conditions. Moreover, we find that the critical photon numbers relevant to ionization are not simply related to the critical photon number $n_{\rm crit}$ of the Jaynes-Cummings Hamiltonian at which the dispersive approximation breaks down~\cite{Blais2004Cavity}. Indeed, while the latter indicates when the qubit-induced nonlinearity becomes important, it is not a good predictor for the onset of ionization. Throughout this work, we use the term ``ionization'' loosely, that is, even if in some instances the measurement-induced transitions do not involve states above the cosine potential well. An alternative term describing the same phenomenology is measurement-induced state transitions~\cite{Khezri2023Measurement,Sank2016Measurement}. However, since the physics at play extends well beyond measurement processes, we have chosen to retain the use of ``ionization'' despite its limitations.

\section{Fully quantum model}\label{sec:Fully quantum model}

\begin{figure}[t!]
\centering
\includegraphics[width=\linewidth]{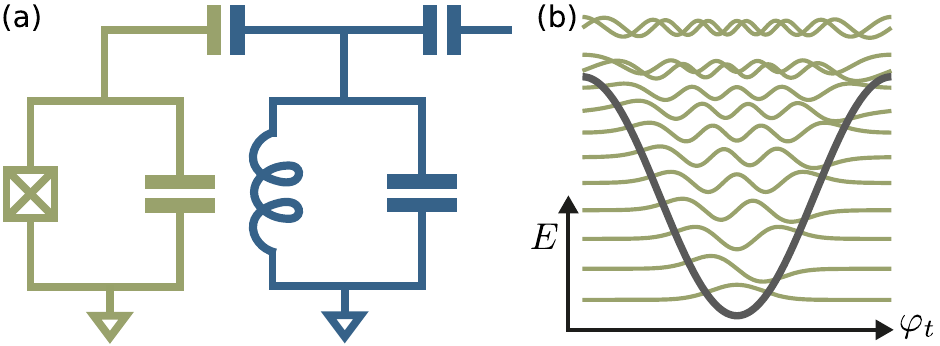}
\caption{(a) Schematic of a transmon qubit capacitively coupled to a driven resonator. (b) Cosine potential well of the transmon with $E_J/E_C = 110$ and $n_g = 0$, together with the first 13 eigenstates represented in the phase basis. There are about nine levels in the well. Transmon states at the bottom of the well are close to eigenstates of the harmonic oscillator, while states above the well are close to charge states, i.e., standing waves in the phase basis.
\label{fig:circuit_cosine_well}}
\end{figure}

\subsection{Model}

As illustrated in \cref{fig:circuit_cosine_well}(a), we consider a transmon qubit capacitively coupled to a resonator described by the Hamiltonian ($\hbar=1$)~\cite{Koch2007ChargeInsensitive, Blais2021Circuit}
\begin{equation}
\begin{split}
 \hat{H}_{tr}
 &=
 \omega_r \ahat^\dagger \ahat + \hat{H}_t
 -i g(\hat{n}_t-n_g)(\ahat-\ahat^\dagger),
\end{split} \label{eq:transmon_resonator_hamiltonian}
\end{equation}
where $\omega_r$ is the bare resonator frequency, $g$ is the transmon-resonator coupling strength, and $\ahat$ is the annihilation operator of the resonator. Moreover, 
\begin{equation}
 \hat{H}_t = 4 E_C (\hat{n}_t-n_g)^2 - E_J \cos \hat{\varphi}_t
\end{equation} is the transmon Hamiltonian with $E_C$ the charging energy, $E_J$ the Josephson energy, and $n_g$ the gate charge. The operators $\hat{\varphi}_t$ and $\hat{n}_t$ are the transmon's canonically conjugate phase and charge operators, respectively. Throughout this work, we assume that the qubit is operated in the transmon regime, $E_J/E_C \gg 1$. Note that we include the often-dropped contribution of the gate charge to the last term of $\hat{H}_{tr}$.

The resonator drive used to measure the qubit takes the form
\begin{align}
 \hat{H}_{d} (t)
 =
 -i
 \varepsilon_d \sin (\omega_d t)
 \left(
 \ahat - \ahat^\dagger
 \right), 
 \label{eq:drive_hamiltonian}
\end{align}
with $\varepsilon_d$ the drive amplitude and $\omega_d$ the drive frequency. Including single-photon resonator losses, the master equation for the transmon-resonator density matrix is then~\cite{Blais2021Circuit}
\begin{align}\label{eq:Full_ME}
 \partial_t \hat{\rho} =
 -i[\hat{H}_{tr} + \hat{H}_{d}(t) , \hat{\rho}] + \kappa \mathcal{D} [\ahat] \hat{\rho},
\end{align}
where $\kappa$ is the resonator decay rate and $\mathcal D[\hat a]\hat \rho = \hat a \hat \rho \hat a^\dag - \{\hat a^\dag \hat a, \hat \rho\}/2$ is the Lindblad dissipator.

In the absence of drive and qubit-resonator coupling, we denote the eigenstates of the Hamiltonian by $|i_t, n_r\rangle$, where $i_t$ labels the bare transmon eigenstates and $n_r$ labels the bare resonator photon number. The bare energy associated with $|i_t, n_r\rangle$ is $E_{i_t} + n_r \omega_r$, and the transition frequencies between different states of the transmon are denoted $\omega_{i_tj_t} \equiv E_{j_t} - E_{i_t}$. With this notation, the bare qubit transition frequency is $\omega_q = E_{1_t} - E_{0_t}$. The matrix elements of the qubit-resonator coupling are proportional to $g_{i_tj_t} = g \langle i_t|\hat{n}_t|j_t \rangle$.

As is usual for quantum information processing in circuit QED, throughout this work we assume that the system is operated in the dispersive limit where $|g/\Delta| \ll 1$, with $\Delta = \omega_q - \omega_r$ the qubit-resonator detuning. In that regime, the computational subspace of the qubit is spanned by the dressed ground state $|\overline{0_t, 0_r}\rangle$ and dressed excited state $|\overline{1_t, 0_r}\rangle$ of $\hat H_{tr}$. Logical operations act on these two dressed states, and it is those states that are resolved in a dispersive qubit measurement. A more precise condition for the dispersive regime to hold is that $\bar n_r/n_{{\rm crit}, i_t} \ll 1$, where $\bar n_r$ is the average resonator photon number and $n_{{\rm crit}, i_t}$ is a photon number at which the transmon-resonator interaction stops being dispersive for state $i_t$ (see \cref{App. Sec.:dispersive_ncrit}). With the above notation, the commonly used Jaynes-Cummings critical photon number is $n_{\rm crit} \equiv n_{{\rm crit}, 0_t} = |\Delta/2g_{0_t 1_t}|^2$~\cite{Blais2004Cavity,Blais2021Circuit}.
As becomes clear below, the breakdown of the dispersive approximation during qubit readout does not necessarily correlate with ionization. 

Unless otherwise stated, throughout this work we fix the transmon parameters to be $E_J/E_C=110$ with $E_C/2\pi=\SI{220}{MHz}$. This corresponds to a qubit transition frequency $\omega_q/2\pi=\SI{6.298}{GHz}$, anharmonicity of magnitude $\alpha/2\pi=\SI{240}{MHz}$, and nine or ten levels in the cosine potential well depending on the value of $n_g$; see \cref{fig:circuit_cosine_well}(b). Moreover, we take the coupling to be $g/2\pi=\SI{120}{MHz}$. Because of the negative sign of the transmon's anharmonicity, there is a striking difference in the onset of ionization depending on the sign of the qubit-resonator detuning $\Delta$~\cite{Khezri2023Measurement,Cohen2023Reminiscence}. In this work, we study both negative and positive detunings. When considering negative detuning, the resonator frequency is $\omega_r/2\pi=\SI{7.5}{GHz}$ ($\Delta/2\pi = \SI{-1.202}{GHz}$, $n_{{\rm crit}, 0_t} = 14$, $n_{{\rm crit}, 1_t} = 10.5$), while for positive detuning we take $\omega_r/2\pi = \SI{5.3}{GHz}$ ($\Delta/2\pi = \SI{0.998}{GHz}$, $n_{{\rm crit}, 0_t} = 9.7$, $n_{{\rm crit}, 1_t} = 2.9$). In \cref{sec: Floquet with just transmon,sec:chaos}, the driven resonator is substituted by a direct drive on the transmon. There, we take $\omega_d/2\pi=\SI{7.515}{GHz}$ at negative detuning and $\omega_{d}/2\pi = \SI{5.267}{GHz}$ at positive detuning. In both cases, the chosen drive frequency is positioned between the two pulled resonator frequencies corresponding to the qubit's ground and excited states~\cite{Blais2021Circuit}. 

Crucially, although we use a fixed set of parameters to introduce the ionization phenomenology, we verified that the conclusions we reached remained general over a wide range of parameters; see, e.g.,~\cref{subsec:Critical_photon_number} and \cref{sec:prediction_experiment}, where we present sweeps over a wide range of resonator frequencies and Josephson energies, respectively. We find that once the sign of the detuning $\Delta$ has been fixed, the phenomenology for a given choice of parameters is easily understood. This can be traced back to the fact that the properties of the transmon relevant to ionization, such as its anharmonicity, the gate charge dispersion of high-energy levels, and the general structure of the charge operator, are roughly independent of the ratio $E_J/E_C$. Furthermore, ignoring quantum fluctuations, a change in the coupling strength $g$ can be understood as a rescaling of the average resonator photon number.

\subsection{Branch analysis}
\label{sec:BA}

To understand how the presence of photons in the resonator leads to transmon ionization, we can attempt to label dressed states $|\overline{i_t, n_r}\rangle$ by considering which is the closest to a given bare transmon state $\ket{i_t}$ at an arbitrary photon number $\ket{n_r}$. Each set of states associated with that particular $i_t$ is then referred to as a branch. While this identification can easily be done at low photon numbers, there is no clear such closest state when the photon number approaches or exceeds $n_{\rm crit}$, in which case the dressed states are highly entangled qubit-resonator states. To build branches that allow us to gain insight into the dynamics of the driven system at high photon numbers, we instead follow Refs.~\cite{Shillito2022Dynamics,Boissonneault2010Improved}. In this approach, we use as our starting point the eigenstates $\{|\lambda \rangle\}$ of $\hat{H}_{tr}$ obtained from numerical diagonalization. For each $i_t$, we identify $|\overline{i_t, 0_r}\rangle$ to be the low-energy eigenstate with the largest overlap with $|i_t, 0_r\rangle$. The branches are built recursively from those starting points and in parallel: Given $|\overline{i_t, n_r}\rangle$, the next eigenstate $|\overline{i_t, n_r+1}\rangle$ added to each branch is the one which maximizes the overlap
\begin{align}\label{eq:C_overlap}
 C_{\overline{i_t, n_r}}(\lambda) 
 \equiv 
 | \langle \lambda| \ahat^\dagger |\overline{i_t, n_r}\rangle |^2.
\end{align}
At each step, $|\lambda\rangle$ is taken from the set of unassigned states. In this way, we obtain branches labeled $B_{i_t}$ corresponding to the set of eigenstates $\{\ket{\overline{i_t,n_r}}\}$ with fixed $i_t$ and with $n_r$ spanning the full resonator Hilbert space. Thus, when the dispersive approximation is valid, each such branch can be pictured as an effective oscillator responding at the pulled resonator frequency associated with the transmon state $i_t$~\cite{Shillito2022Dynamics}. 

\begin{figure}[t!]
 \centering
 \includegraphics[width=\linewidth]{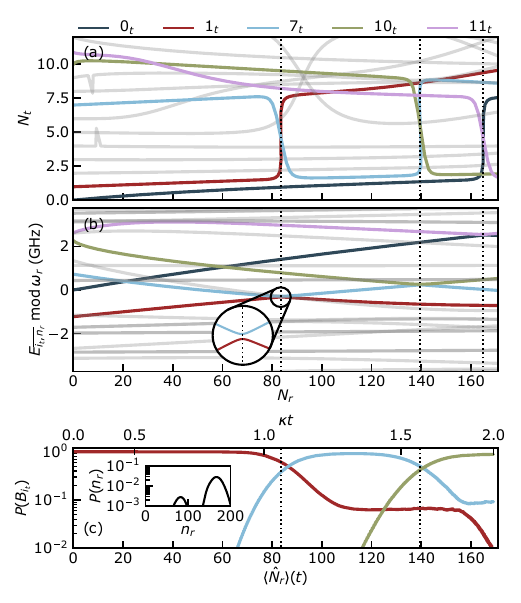}
 \caption{
 {\bf Negative detuning.} (a) Average transmon population $N_t$ of branch eigenstates and (b) modular eigenenergies of branch eigenstates as a function of the average resonator photon number $N_r$ of the eigenstates. All results in (a,b) are obtained from full diagonalization of the undriven Hamiltonian $\hat H_{tr}$ for a negative detuning $\Delta/2\pi = -\SI{1.202}{GHz}$. In (a,b), branches $B_{0_t}$, $B_{1_t}$, $B_{7_t}$, $B_{10_t}$, and $B_{11_t}$ are highlighted in color (see the legend), and the others are gray. In (b), energies are measured from the qubit ground state energy and then folded into the interval $[-\omega_r/2, \omega_r/2]$. Branch population swapping occurs at avoided crossings in the modular energy spectrum (dotted vertical lines and inset). (c) Dynamics of the driven transmon-resonator system when initializing in the excited state $\ket{\overline{1_t,0_r}}$. When the resonator population reaches $\langle\ahat^\dag \ahat \rangle\approx 84$, corresponding to the $B_{1_t}-B_{7_t}$ swapping observed in (a), the population of the state is mostly transferred to branch $B_{7_t}$, but a significant part of the population (approximately $7\%$) stays in branch $B_{1_t}$, indicating ionization. At $\langle\ahat^\dag \ahat \rangle\approx 140$, the population is then mostly transferred to $B_{10_t}$, corresponding to the $B_{7_t}-B_{10_t}$ swapping in (a), while residual population in $B_{7_t}$ is ionized. The inset in (c) shows in log scale the bimodal Fock-state distribution of the resonator at the final time $\kappa t = 2$. For dynamics simulations, we use $\varepsilon_d/2\pi=\SI{180}{MHz}$, $\omega_d/2\pi=\SI{7.515}{GHz}$, and $\kappa/2\pi=\SI{7.95}{MHz}$ (set to match the full $\chi$ shift at zero photons). The full-time dynamics are performed using quantum trajectories~\cite{Carmichael_Textbook_2} with 16 transmon states, 300 resonator states, 200 trajectories, and time evolving up to $\kappa t = 2$. The timescale of the simulated dynamics is shown on the secondary axis above (c).
 }
 \label{fig: fully quantum negative detuning}
\end{figure}

The character of the branch eigenstates can be investigated by tracking their average transmon population $N_{t}(i_t,n_r) \equiv \sum_{j_t, m_r} j_t|\langle j_t, m_r |\overline{i_t,n_r}\rangle|^2$ and their average resonator population $N_{r}(i_t,n_r) \equiv \langle \overline{i_t, n_r} | \ahat^\dagger \ahat |\overline{i_t, n_r} \rangle$ for every eigenstate $\ket{\overline{i_t, n_r}}$. Note that whenever we refer to $N_t$ and $N_r$ of a specific branch, we are implicitly referring to the $N_t$ and $N_r$ of the eigenstates which make up that branch. From this point onward, we stop writing the explicit indices $(i_t,n_r)$ to lighten the notation. At very low photon numbers, the dispersive approximation holds, and we expect the states within a given branch $B_{i_t}$ to have a transmon population $N_t$ close to the bare value $i_t$. However, plotting the average transmon and resonator populations parametrically for each branch reveals that this is not always the case; see \cref{fig: fully quantum negative detuning}(a). Instead, drastic changes in the average transmon population $N_t$ are sometimes observed at specific branch-dependent values of $N_r$, such as a ``swapping" of the branches in which two branches exchange their values of $N_t$. These features are a signature of strong hybridization between transmon states and indicate a complete breakdown of the dispersive approximation. When swapping occurs between a branch linked to computational states and a branch that is near the top of the cosine potential, there is a significant change in the character of the former. It shifts from being predominantly made up of low-energy, localized bare transmon states to being composed largely of high-energy, chargelike bare transmon states. Thus, we generically refer to this process as ionization. 

In Refs.~\cite{Shillito2022Dynamics,Shillito2023Thesis}, these drastic changes were shown to determine dynamical properties of the system in the presence of a readout drive populating the resonator. Thus, the branch analysis serves as a diagnosis of the onset of measurement-induced ionization in the transmon. Because of computational limitations, however, these works mostly focused on a single parameter set. As a result, the difference in the underlying physics at negative and positive detunings was not explored in great detail, nor was the impact of gate charge studied. Below, we describe the physical mechanisms relevant for negative transmon-resonator detuning, followed by the case where this detuning is positive. Finally, we quantify the effect of gate charge on the transitions.

\subsection{Negative detuning \texorpdfstring{$\omega_q < \omega_r$}
 {\unichar{"03C9}\unichar{"1D45E} < \unichar{"03C9}\unichar{"1D45F}}}
\label{subsec: fully quantum negative detuning}

We consider the case where the qubit frequency is below the resonator frequency, $\Delta = \omega_q - \omega_r < 0$. To proceed, we note that by taking matrix elements of the commutation relation $[\hat{H}_{tr}, \ahat^\dagger] = \omega_r \ahat^\dagger - i g (\hat{n}_t-n_g)$ and rearranging, \cref{eq:C_overlap} can be rewritten as~\cite{Cohen2023Reminiscence}
\begin{align}\label{eq:C_new_form}
 C_{\overline{i_t, n_r}}(\lambda) 
 =
 g^2
 \frac{
 |\langle \lambda | \hat{n}_t |\overline{i_t, n_r}\rangle|^2
 }
 {
 (
 E_\lambda - E_{\overline{i_t, n_r}} - \omega_r
 )^2
 }.
\end{align}
The denominator of this expression conveys that when adding a photon to a branch eigenstate makes it nearly degenerate with another eigenstate, dramatic effects such as branch swapping can be expected.
%makes two nearby eigenstates within a branch nearly degenerate
But when do we expect dressed states of $\hat{H}_{tr}$ to be nearly resonant?

Perturbation theory provides an intuitive answer to this question. Using the usual Schrieffer-Wolff transformation $e^{-\hat{S}}$ with an appropriately chosen $\hat{S}$, one can approximately diagonalize $\hat{H}_{tr}$ as~\cite{Koch2007ChargeInsensitive}
\begin{align} \label{eq:SW_H_0}
e^{\hat{S}}
 \hat{H}_{tr}
 e^{-\hat{S}}
 &\approx
 \sum_{i_t = 0 }^\infty
 \left[
 E_{i_t} + \Lambda_{i_t}
 +
 (\omega_r + \chi_{i_t}) \ahat^\dagger \ahat
 \right]
 |i_t\rangle\langle i_t |,
\end{align}
where $\Lambda_{i_t}$ are the Lamb shifts and
\begin{align}\label{eq:chi_shifts}
 \chi_{i_t}
 =
 \sum_{j_t = 0}^{\infty}
 |g_{i_t j_t}|^2
 \left(
 \frac{1}{\omega_r - \omega_{i_t j_t}}
 -
 \frac{1}{\omega_r + \omega_{i_t j_t}}
 \right)
\end{align}
are the dispersive $\chi$ shifts obtained here without the usual rotating-wave approximation (RWA) on the transmon-resonator coupling term of \cref{eq:transmon_resonator_hamiltonian}~\cite{Blais2021Circuit}. Importantly, because of the transmon's negative anharmonicity, at negative detuning the transitions between nearest-neighbor higher-excited states are more off resonant with the resonator, e.g.,~$|\Delta - \alpha| > |\Delta|$. This trend continues for all transmon eigenstates $i_t$ within the cosine potential and ensures the validity of perturbation theory at low photon numbers. Given the small change $\chi_{i_t}$ to the dispersion of the resonator, at low photon numbers we then have a direct mapping between the bare states $|i_t, n_r\rangle$ and dressed states $|\overline{i_t,n_r}\rangle \approx e^{-\hat{S}}|i_t, n_r\rangle \equiv |\dashover{i_t,n_r}\rangle$. 

To understand how this standard analysis helps explain the phenomenon of branch swapping, it is instructive to consider the eigenvalues of $\hat{H}_{tr}$ modulo $\omega_r$. This modular spectrum is plotted versus resonator photon number in \cref{fig: fully quantum negative detuning}(b). At low photon numbers, the slope of the modular branch energy versus $N_r$ for states in the well is $\chi_{i_t}$, which is positive for states deep in the well and negative for states near the top of the well. This change in sign can be understood as a competition between the two terms in \cref{eq:chi_shifts}, corresponding to virtual photon absorption and emission, respectively. The most relevant transitions for absorption are those for which $j_t > i_t$, which have a larger matrix element. By contrast, the most relevant transitions for emission are those for which $j_t < i_t$, which are less detuned due to the negative anharmonicity of the transmon. Indeed, treating the transmon as a Kerr-nonlinear oscillator with $\alpha \approx -E_C$, performing the RWA, and setting $n_{\rm zpf}$ as the zero-point fluctuations of the charge, the leading absorption and emission terms are found to be
\begin{align}
 \chi_{i_t}
 \approx
 g^2 n^2_{\rm zpf}
 \left(
 \frac{i_t+1}{-\Delta + i_t E_C}
 -
 \frac{i_t}{-\Delta+(i_t-1) E_C}
 \right),
\end{align}
thereby correctly predicting the general trend of $\chi_{i_t}$ decreasing with increasing $i_t$ for negative detuning. Crucially, however, this approximation incorrectly predicts the same sign of $\chi_{i_t}$ for all transmon states, highlighting the importance of keeping the full cosine potential.

At large enough photon numbers, these opposite ac-Stark shifts can lead to near collisions of the modular energies between states at the bottom and near the top of the well, corresponding to near degeneracies $E_{\lambda} - E_{\lambda'} \approx 0 \mod \omega_r$ of the eigenenergies. Nonperturbative corrections to the eigenstates -- which arise due to terms typically dropped in the derivation of \cref{eq:SW_H_0} -- turn these near collisions into avoided crossings (dotted vertical lines); see \cref{App. Sec.: Non-perturbative corrections}. These avoided crossings in the modular spectrum correspond to multiphoton resonances where $k \sim |E_{\lambda} - E_{\lambda'}|/\omega_r$ photons are absorbed by the transmon. Accordingly, the states before and after an avoided crossing swap character: If the energies of the eigenstates of branches $B_{i_t}$ and $B_{j_t}$ collide modulo $\omega_r$, we then have 
$|\overline{i_t, n_r}\rangle \approx |\dashover{i_t, n_r}\rangle$, $|\overline{j_t, n_r}\rangle \approx |\dashover{j_t, n_r}\rangle$ before the avoided crossing and 
$|\overline{i_t, n_r}\rangle \approx |\dashover{j_t, n_r+k}\rangle$, $|\overline{j_t, n_r}\rangle \approx |\dashover{i_t, n_r-k}\rangle$ after the avoided crossing. We emphasize that, since states near the top of the well break the dipolelike selection rules of the Kerr nonlinearity approximation, it is not required that multiphoton transitions satisfy the condition $|i_t - j_t| = k$. Thus, preserving the full cosine potential well of the transmon is necessary to accurately predict the strength of the avoided crossing. 

These qualitative predictions capture the main features observed in \cref{fig: fully quantum negative detuning}(a,b). As expected, at low photon numbers the modular energies are observed to vary linearly with photon number. The slope is positive for branches corresponding to states deep in the well, such as $B_{1_t}$, while the slope is negative for branches corresponding to states near the top, such as $B_{7_t}$. An avoided crossing is observed when these two branches eventually meet at $N_r \approx 84$; see the inset in \cref{fig: fully quantum negative detuning}(b). This is precisely where the branches swap characters; see \cref{fig: fully quantum negative detuning}(a). The same phenomenology holds for the ground-state branch $B_{0_t}$, which swaps with the chargelike branch $B_{11_t}$ after an avoided crossing at a higher photon number $N_r \approx 165$. An avoided crossing between $B_{7_t}$ and $B_{10_t}$ is also observed at $N_r \approx 140$. Importantly, not all near degeneracies lead to swapping, such as when $B_{0_t}$ and $B_{7_t}$ cross at around $N_r \approx 23$. This is simply a consequence of the magnitude of the relevant matrix elements: At this low photon number, it is not possible to efficiently couple the dressed $0_t$ and $7_t$ states of the transmon via the qubit-resonator coupling; see \cref{App. Sec.: Non-perturbative corrections} for details. Note that ionization here happens at a photon number that is much larger than the Jaynes-Cummings critical photon number.

We stress that, although the above leading-order perturbative approach used to interpret \cref{fig: fully quantum negative detuning} is qualitatively correct, to obtain quantitative predictions one must resort to full numerics. For instance, at large enough photon numbers, the self-Kerr of the dressed resonator becomes relevant and must be taken into account when trying to predict the positions of the avoided crossings. In addition, ionization of the ground- and excited-state branches can occur with chargelike states. Indeed, given that $\hat{n}_t$ can efficiently couple states near the top of the well to each other and even to chargelike states outside the cosine potential via a single photon~\cite{Cohen2023Reminiscence}, the modular spectrum of these highly excited states can vary wildly as a function of photon number, thus bringing them in and out of resonance with the computational states. The strong charge dispersion of the transmon states near and above the top of the cosine potential well then implies that $n_g$ plays a crucial role in determining at which photon number ionization occurs~\cite{Cohen2023Reminiscence,Khezri2023Measurement}. See \cref{subsec:gate_charge} for more details on these considerations. 

It is important to emphasize that, in the language of the branch analysis, ionization occurs when the population remains in the ground- or excited-state branch after the swapping; consider e.g.~the red line in \cref{fig: fully quantum negative detuning}(a), which jumps from $N_t \sim 1$ to $N_t \sim 7$ at the avoided crossing. This corresponds to an adiabatic transition to a highly excited transmon state. By contrast, purely diabaticlike transitions between branches do not cause ionization. In practice, the character of the transition can be mixed between these two types of processes. As an illustration, by starting in $|\overline{1_t, 0_r}\rangle$, time evolving via \cref{eq:Full_ME}, and plotting the average population of the branches $P_{B_{i_t}}(t) \equiv \langle \sum_{n_r}|\overline{i_t, n_r}\rangle \langle \overline{i_t, n_r} | \rangle$ as in \cref{fig: fully quantum negative detuning} (c), we see a nearly full diabatic population transfer from $B_{1_t}$ to $B_{7_t}$ at $\langle \ahat^\dagger \ahat \rangle \approx 84 $, with the bare transmon population (not shown) rising only to $N_t \approx 2.25$. The approximately 7\% of the population that remains in $B_{1_t}$ after the swapping roughly translates to an equally large probability of ionization. Ionizing by remaining in the ground- or excited-state branch after the crossing implies a change in the resonator pull, which will in this instance be closer to $\chi_{7_t}$ than to $\chi_{1_t}$. As discussed in Ref.~\cite{Shillito2022Dynamics}, this, in turn, implies that the resonator undergoes a different phase-space trajectory, leading to the bimodal distribution of Fock states seen in the inset in \cref{fig: fully quantum negative detuning}(c). 

The importance of multiphoton resonances as a mechanism causing drive-induced transitions in the transmon has previously been discussed in the literature. \textcite{Sank2016Measurement} show that resonances between different excitation-conserving subspaces of the Jaynes-Cummings Hamiltonian mediated by several photons can lead to transmon transitions. The theoretical model developed in that work, which identifies multiphoton resonances using an approach related to the branch analysis of \cref{fig: fully quantum negative detuning}, was able to explain experimentally observed features. Interestingly, ionization was observed to occur at a photon number significantly larger than the Jaynes-Cummings critical photon number, as is expected from the above discussion. Moreover, that experiment also shows that while readout-induced transitions occur at specific photon numbers, it is possible to operate the readout at still larger photon numbers. This highlights that, as discussed above, resonances that are traversed diabatically do not lead to ionization. The importance of these resonances was also recognized in the context of the resonator-induced phase gate in Ref.~\cite{MalekakhlaghRIPLeakage}.

Moreover, \textcite{Xiao2023Diagrammatic} use a diagrammatic method to compute effective Hamiltonians for a driven transmon in the absence of a resonator. This approach allows them to identify the multiphoton processes that are responsible for drive-induced transmon transitions. In the next section, we show that the model of a directly driven transmon can capture ionization due to a readout drive. The branch analysis brings these pictures together and provides a simple way to predict where resonances occur. It also captures resonances that cannot be obtained via perturbative arguments. This occurs, for instance, when invoking the RWA for higher-energy states of the cosine potential, which would lead one to incorrectly predict the photon number at which ionization occurs; see \cref{subsec:gate_charge}.

We can now succinctly summarize the three key ingredients leading to ionization of the qubit ground and excited states at negative detuning.

\begin{itemize}

\item At low photon numbers, $\chi_{i_t}$ is positive for states deep in the well, leading to an increase in the energy with photon number. In contrast, transmon states near the top of the cosine potential are pushed down in energy with increasing photon number. Chargelike states above the well can also be pushed down in energy in a manner which is highly sensitive to the gate charge.

\item These opposite behaviors result in near-degeneracies and, thus, in avoided crossings between the two sets of states at some large-enough photon number, leading to swapping of the transmon branches.
 
\item The resulting hybridization of the transmon's computational states with states closer to the top of the well or above the well leads to transmon ionization at specific resonator photon numbers.
\end{itemize}

\subsection{Positive detuning \texorpdfstring{$\omega_q > \omega_r$}
 {\unichar{"03C9}\unichar{"1D45E} > \unichar{"03C9}\unichar{"1D45F}}}
\label{subsec: fully quantum positive detuning}

\begin{figure}[t!]
 \centering
 \includegraphics[width=\linewidth]{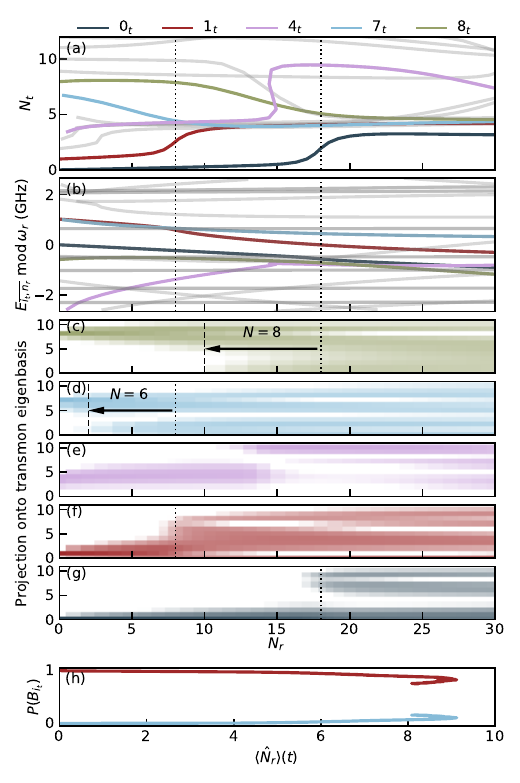}
 \caption{{\bf Positive detuning.} (a) Average transmon population $N_t$ of branch eigenstates, (b) modular eigenenergies of branch eigenstates, and (c-g) projection of the branch eigenstates on the transmon eigenbasis as a function of the average resonator photon number $N_r$ of the eigenstates. All results in (a-g) are obtained from full diagonalization of the undriven Hamiltonian $\hat H_{tr}$ for a positive detuning $\Delta/2\pi = \SI{0.998}{GHz}$. In (a,b), branches $B_{0_t}$, $B_{1_t}$, $B_{4_t}$, $B_{7_t}$, and $B_{8_t}$ are highlighted in color (see the legend), and the others are gray. In (b), energies are measured from the qubit ground state energy and then folded into the interval $[-\omega_r/2, \omega_r/2]$. In (c-g), the projections of $8_t$, $7_t$, $4_t$, $1_t$, and $0_t$ are plotted on a logarithmic color bar scale with the same color coding. Branch population bunching is observed about state $i_t^\star = 4$. The dotted vertical lines indicate the photon numbers at which the population bunching in $B_{0_t}$ and $B_{1_t}$ dramatically increases. The dashed vertical lines are offset from the dotted lines by $N$ photons, with $N$ being the number of photons needed to energetically connect the pairs of states that hybridize during bunching. (h) Dynamically simulated populations of branches $B_{1_t}$ (red line) and $B_{7_t}$ (light blue line) of the driven transmon-resonator system when initializing in the excited state $\ket{\overline{1_t,0_r}}$. Dynamics simulations are done similarly to \cref{fig: fully quantum negative detuning}(c) but with $\varepsilon_d/2\pi=\SI{93}{MHz}$, $\omega_d/2\pi=\SI{5.267}{GHz}$, $\kappa/2\pi=\SI{11.98}{MHz}$, and 120 resonator states.}
 \label{fig:pos_branch_analysis}
\end{figure}

In contrast to negative detuning where branch swapping is observed, at moderate positive qubit-resonator detuning we instead observe that the transmon populations of the branches within the cosine potential well coalesce as the resonator photon number is increased; see \cref{fig:pos_branch_analysis}(a). Much of the phenomenology associated with this observation stems from the fact that the transmon's negative anharmonicity leads to transitions between neighboring states in the well becoming more resonant with the resonator frequency when climbing the cosine potential. As a result, because of the small anharmonicity of the transmon and for typical large values of $g$ in circuit QED, generically one of the transmon transitions within the well is such that $|E_{i_t+1} - E_{i_t} - \omega_r| \sim g$. For that transition, the dispersive approximation breaks down even at zero resonator photon population. In the Kerr-nonlinear oscillator approximation for the transmon, where $\omega_{i_t,i_t+1} \sim \omega_q- i_t \alpha$, this one-photon resonance occurs for the transmon level $i_t^\star \sim \mathrm{round}[(\omega_q - \omega_r)/\alpha]$~\cite{Khezri2023Measurement}. 

For example, with the parameters in \cref{fig:pos_branch_analysis} we have that
\begin{equation}\label{eq:pos_simple_inverted_well}
\begin{cases}
 E_{i_t+1} - E_{i_t} > \omega_r & \text{if $i_t \leq 3$}, \\
 E_{i_t+1} - E_{i_t} \approx \omega_r & \text{if $i_t = 4$},\\
 E_{i_t+1} - E_{i_t} < \omega_r & \text{if $i_t \geq 5$},
\end{cases}
\end{equation}
so that $i_t^\star = 4$ is the state for which the dispersive approximation first fails, $E_{i^\star_t+1} - E_{i^\star_t} - \omega_r \sim g$. This is confirmed by examining the modular energies of the branches eigenstates; see \cref{fig:pos_branch_analysis}(b). Unlike the prediction made by the dispersive approximation in \cref{eq:SW_H_0}, $E_{\overline{4_t, n_r}}$ does not vary linearly at small photon numbers (pink line), indicating a strong mixing of the bare state $|4_t, n_r\rangle$ with other neighboring bare transmon states. This strong hybridization results in the delocalization of $4_t$ at very low photon numbers, as shown in \cref{fig:pos_branch_analysis}(e). Increasing the photon number, hybridization of $i_t^\star$ with neighboring states leads to level repulsion and, in turn, to collisions with other branches. 

At the moderate positive detuning considered here, this hybridization is further accentuated by near resonances (modulo $\omega_r$) between pairs of states in the well. For the parameters in \cref{fig:pos_branch_analysis}, this is the case of the pair of states $1_t$ and $7_t$, for which $E_{7_t}-E_{1_t} - 6 \omega_r \sim g$, and of $0_t$ and $8_t$, for which $E_{8_t}-E_{0_t} - 8 \omega_r \sim g$. These transitions require multiple intermediate virtual transitions and are, thus, suppressed at very low photon numbers; see in \cref{fig:pos_branch_analysis}(b) the almost resonant modular energies of the eigenstates of branches $B_{1_t}$ and $B_{7_t}$ running parallel up to $N_r \approx 7$. However, because the coupling matrix element is proportional to $g\sqrt{n_r}$, their hybridization becomes possible at increased photon numbers; see \cref{fig:pos_branch_analysis}(c,d,f,g). The early breakdown of the dispersive approximation for $i_t^\star$, followed by the activation of near resonances, is the cause of the observed bunching of the branch population at $N_t\sim i_t^\star$. This process starts with states closest to $i_t^\star$ and, pair by pair, is followed by states connected by a multiphoton transition and that are symmetric about $i_t^\star$. This phenomenology explains why branch bunching and the associated ionization typically occur at a lower photon number for the transmon's excited state than for its ground state; see \cref{fig:pos_branch_analysis}(a). It also explains why, at moderate detuning, ionization typically occurs at much lower photon numbers for positive detuning than for negative detuning. 

\Cref{fig:pos_branch_analysis}(h) shows the average population of the branches when starting in $|\overline{1_t, 0_r}\rangle$ and evolving using \cref{eq:Full_ME}. Around $\langle\hat{N}_r\rangle(t)\approx 7$, which coincides with branches $B_{1_t}$ and $B_{7_t}$ bunching around $N_r\approx 7$ in \cref{fig:pos_branch_analysis}(a), a fraction of the population is transferred from branch $B_{1_t}$ to $B_{7_t}$. However, even at the relatively high speed at which the resonance is traversed, most of the population remains in branch $1_t$ due to the large width of the avoided crossing, resulting in near-complete ionization. The resonator population drops after the crossing due to the drive becoming significantly off resonant with the pulled resonator frequency associated with the ionized state.

\begin{figure}[t]
 \centering
 \includegraphics[width=\linewidth]{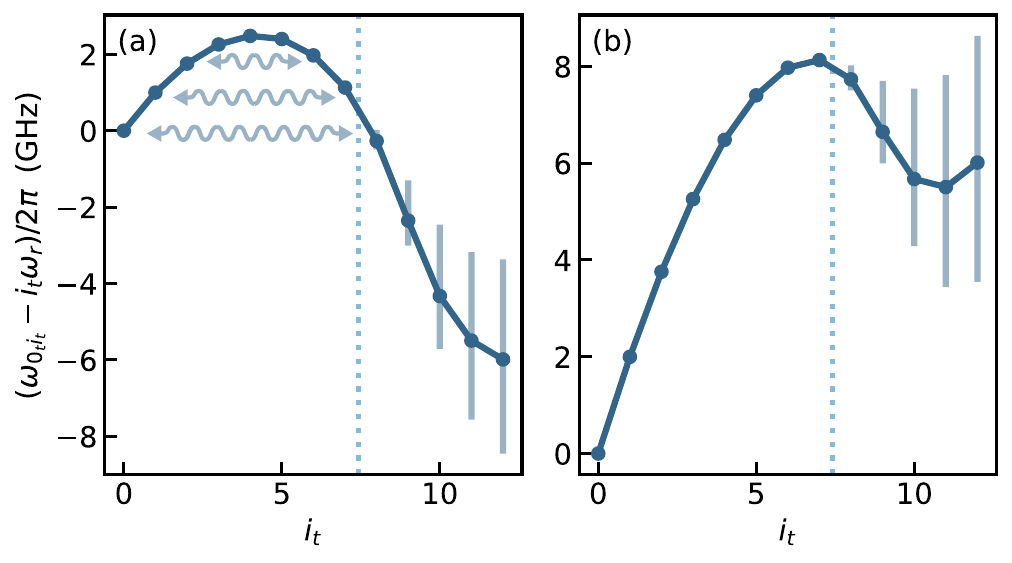}
 \caption{{\bf Positive detuning.} Bare energies $\omega_{0_ti_t} - i_t\omega_r$ of the sites $\ket{i_t,N-i_t}$ in the $N$-excitation subspace of $\hat H_0$ in the RWA, here illustrated for $N=12$. (a) Parameters are as in \cref{fig:pos_branch_analysis}, but with $g=0$ and $n_g = 0.25$. The site $\ket{4_t,N-4_t}$ sits at the top of the inverted parabola. The arrows represent multiphoton processes connecting near-degenerate sites at the two extremities of the lattice. The solid vertical lines represent the charge dispersion of level $i_t$. The dotted vertical line at $2E_J/\omega_p \sim 7.4$ indicates a lower bound for the number of states in the cosine potential well. Here, $\omega_p = \sqrt{8 E_C E_J}$ is the plasma frequency. (b) The same parameters except for a larger positive detuning $\Delta/2\pi = \SI{1.998}{GHz}$ ($\omega_r/2\pi = \SI{4.3}{GHz}$).}
 \label{fig:pos_branch_analysis_1Dlattice}
\end{figure}

In Ref.~\cite{Khezri2023Measurement}, this physics was shown to be captured within a semiclassical model in the RWA. In the fully quantum model we use here, this can be understood by projecting the qubit-resonator Hamiltonian $\hat H_0$ in \cref{eq:transmon_resonator_hamiltonian} on a subspace with $N$ excitations. After performing a RWA, dropping $n_g$ from the coupling term, and moving to a frame rotating at $\omega_r$ for the qubit and the resonator, this leads to
\begin{equation}\label{eq:1D_lattice}
 \begin{split}
  \hat H_0^N &= \sum_{i_t+n_r = N} (E_{i_t} - i_t\omega_r) \ket{i_t,n_r}\bra{i_t,n_r} \\
  &- i\sum_{i_t+n_r = N} \left(g_{i_t+1,i_t}\sqrt{n_r} 
  \ket{i_t+1,n_r-1}\bra{i_t,n_r} + \mathrm{H.c.}\right)
 \end{split}
\end{equation}
This is analogous to a 1D lattice with nearest-neighbor hopping amplitudes $g_{i_t+1,i_t}\sqrt{n_r}$. As expected from \cref{eq:pos_simple_inverted_well}, the bare energies of the sites $\ket{i_t,n_r}$ on this lattice form an inverted parabola with $\ket{i_t^\star,N-i_t^\star}$ at the top; see \cref{fig:pos_branch_analysis_1Dlattice}(a)~\cite{Khezri2023Measurement}. For the parameters in \cref{fig:pos_branch_analysis}, the states $\ket{3_t,N-3_t}$, $\ket{4_t,N-4_t}$, and $\ket{5_t,N-5_t}$ near the top of the parabola are near degenerate and are connected by a single lattice hop. Therefore, they rapidly hybridize, leading to the almost immediate state bunching seen at low photon numbers in \cref{fig:pos_branch_analysis}(a). Hybridization between the other pairs of near-degenerate states located on each side of the lattice becomes relevant once the photon number has increased enough to make the hopping amplitude sufficiently large to allow virtual transitions between those two remote states. At that point, the wave functions of that pair of states become delocalized along the 1D lattice, and they join the layer of bunched states. This occurs pair by pair until the photon number is large enough for the ground state sitting at the bottom of the cosine potential well to finally enter the bunching layer together with its partner state; see \cref{fig:pos_branch_analysis}(a).

\begin{figure}[t]
 \centering
 \includegraphics[width=\linewidth]{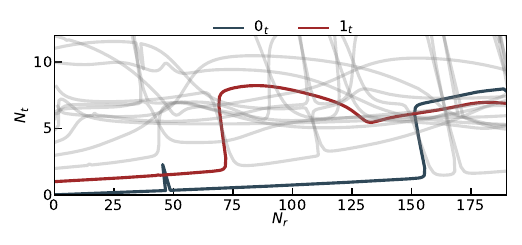}
 \caption{{\bf Positive detuning.} Average transmon population $N_t$ of branch eigenstates as a function of the average resonator photon number $N_r$ of the eigenstates for a larger positive detuning $\Delta/2\pi = \SI{1.998}{GHz}$, corresponding to $i_t^\star = 8$. All other parameters are the same as in \cref{fig: fully quantum negative detuning}. Branches $B_{0_t}$ and $B_{1_t}$ are highlighted in color (see the legend), and the others are gray.
 } 
 \label{fig:pos_branch_analysis_large_detuning}
\end{figure}

This 1D model, however, is not sufficient to explain the situation at larger positive detuning; compare \cref{fig:pos_branch_analysis}(a) to \cref{fig:pos_branch_analysis_large_detuning} obtained for $\Delta/2\pi = \SI{1.998}{GHz}$ ($\omega_r/2\pi = \SI{4.3}{GHz}$). The discrepancy is easily understood. Indeed, because the index $i_t^\star$ of the transmon level sitting at the top of the inverted parabola increases with the detuning, at large detuning states $i_t < i_t^\star$ no longer have quasidegenerate partners on the lattice; see \cref{fig:pos_branch_analysis_1Dlattice}(b). The absence of these multiphoton resonances tends to push ionization to larger photon numbers. 

Moreover, for a small resonator frequency compared to the qubit frequency, more than $k$ photons may be needed to connect a state $|i_t, n_r \rangle$ associated with a transmon state deep in the well to a state $|i_t + k, n_r - k\rangle$ closer to the top of the well. Therefore, considering only the projection on a subspace with a fixed number of excitations is no longer sufficient, signaling the breakdown of the RWA. More generally, \cref{eq:1D_lattice} does not capture the full matrix elements of the transmon charge operator when states near or above the top of the cosine potential well are involved (see below). 

As a result, while a bunching layer can still exist, at large positive detuning branch swapping associated with multiphoton transitions of the type discussed in \cref{subsec: fully quantum negative detuning} is also observed. This is shown in \cref{fig:pos_branch_analysis_large_detuning}, where this occurs for both the ground- and excited-state branches. At large photon numbers, the increase with $\sqrt{n_r}$ of the coupling matrix elements is eventually sufficient to overcome the large energy separations. This leads to significant hybridization, causing the branches to merge with the bunching layer.

To summarize this section, the key ingredients leading to ionization at positive detuning are as follows.
\begin{itemize}
\item Because of the transmon's negative anharmonicity, there generically exists a transmon state $i_t^\star$ for which the dispersive approximation breaks down at a very low photon number. This results in rapid hybridization of the transmon states and to branch bunching close to $N_t \sim i_t^\star$. This branch bunching is, in turn, a signature of ionization.
\item Branch bunching is precipitated by pairwise multiphoton resonances that become relevant as the photon number increases. This is captured within the RWA~\cite{Khezri2023Measurement}.
\item At larger positive detuning, the RWA fails and weaker processes become dominant. As a result, multiphoton resonances involving the computational states are less likely. Generally, the photon number threshold for ionization increases with the detuning.
\end{itemize}

\subsection{Gate charge dependence}
\label{subsec:gate_charge}

An important feature of the transmon is the exponentially small charge dispersion of its computational states with increasing $E_J/E_C$~\cite{Koch2007ChargeInsensitive}. This charge dispersion, however, rapidly increases for states near and above the top of the cosine potential well; see the solid vertical lines in \cref{fig:pos_branch_analysis_1Dlattice}. If the ionizing multiphoton resonances hybridize the computational states with these charge-sensitive states, the ionization threshold will depend on gate charge. This effect is magnified by the fact that single-photon resonances between charge-sensitive states near the top of the well can occur at specific values of the gate charge. The modular energies of these single-photon-hybridized states vary wildly with photon number, pushing them in and out of resonance with the computational states as gate charge is varied. Single-photon resonances at the top of the well are facilitated by the fact that matrix elements such as $\langle i_t|\hat{n}_t|i_t+ 3_t\rangle$, which are essentially zero for computational states, can be large near the top of the well, thus opening up more resonance pathways. When moving away from the symmetry points $n_g = 0$ and 0.5, other matrix elements such as $\langle i_t|\hat{n}_t|i_t+ 2_t\rangle$ also become large. For all these reasons, both the large charge dispersion and the influence of the gate charge on the matrix elements significantly impact the photon number at which ionization of the transmon computational states occurs, even deep in the transmon regime.

\begin{figure}[t!]
 \centering
 \includegraphics[width=\linewidth]{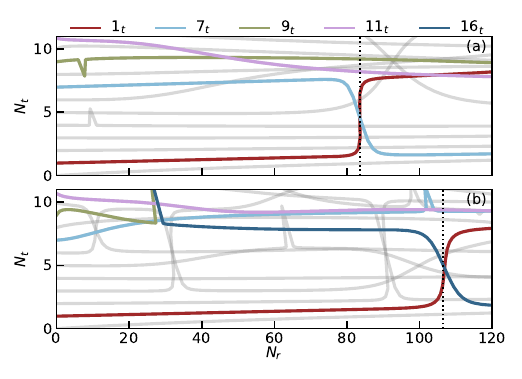}
 \caption{{\bf Negative detuning.} Average transmon population $N_t$ of branch eigenstates as a function of the average resonator photon number $N_r$ of the eigenstates for two different gates charge values (a) $n_g=0$ and (b) $n_g=0.3$. All other parameters are the same as in \cref{fig: fully quantum negative detuning}. Branches $B_{1_t}$, $B_{7_t}$, $B_{9_t}$, $B_{11_t}$, and $B_{16_t}$ are highlighted in color (see the legend), and the others are gray. Ionization of the excited state is pushed back from $N_r\approx 84$ at $n_g=0$ to $N_r\approx 107$ at $n_g=0.3$.}
 \label{fig:resonance_gate_charge}
\end{figure}

This is illustrated in \cref{fig:resonance_gate_charge}, which shows the transmon branch population versus resonator photon number at negative detuning and for two values of the gate charge. For $n_g=0$ [\cref{fig:resonance_gate_charge}(a)], branches $B_{7_t}$ and $B_{1_t}$ form an avoided crossing at around $N_r\approx 84$, resulting in the observed branch swapping. In contrast, for $n_g=0.3$ [\cref{fig:resonance_gate_charge}(b)], because the charge operator can now connect states of the same parity, branch $B_{11_t}$ hybridizes early with $B_{9_t}$ and, in turn, branch $B_{9_t}$ hybridizes with $B_{7_t}$ and then with $B_{16_t}$. The result is a delayed ionization of $B_{1_t}$ through hybridization with $B_{16_t}$ at around $N_r\approx 107$. Dependence on gate charge is observed at positive detuning for similar reasons. The importance of gate charge on the driven transmon was also pointed out in Refs.~\cite{Cohen2023Reminiscence,Khezri2023Measurement}. We note that this gate charge dependence highlights the importance of preserving the full cosine potential of the transmon to study ionization. Indeed, expanding the potential to arbitrary large order allows one to gauge away the gate charge, eliminating gate charge dependence of the eigenspectrum~\cite{Koch2007ChargeInsensitive}.

\subsection{Critical photon number} 
\label{subsec:Critical_photon_number}

\begin{figure}[t]
 \centering
 \includegraphics[width=\linewidth]{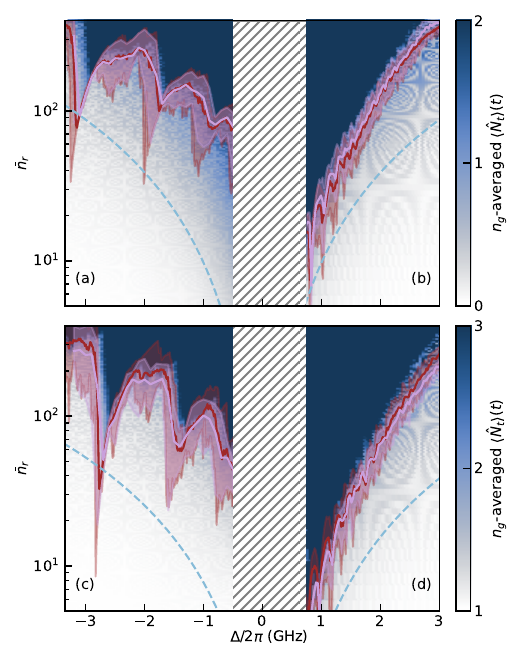}
 \caption{Gate-charge-averaged critical photon number versus detuning extracted from the fully quantum model (pink) and from Floquet analysis (red). The top corresponds to the ground state $\bar{n}_{{\rm crit}, 0_t}^q$ and $\bar{n}_{{\rm crit}, 0_t}^f$ for (a) negative detuning and (b) positive detuning. The bottom corresponds to the excited state $\bar{n}_{{\rm crit}, 1_t}^q$ and $\bar{n}_{{\rm crit}, 1_t}^f$ for (c) negative detuning and (d) positive detuning. The detuning is varied by changing the resonator frequency. Shaded areas correspond to the distribution of $n_{{\rm crit}, i_t}^{q}$ and $n_{{\rm crit}, i_t}^{f}$ between the tenth and 90th percentiles over 100 realizations of $n_g$ for both models. The dashed blue lines are the Jaynes-Cummings-like critical photon numbers $n_{{\rm crit}, i_t}$ defined in \cref{App. Sec.:dispersive_ncrit} for each state. The colored background shows the transmon population extracted from the gate-charge-averaged dynamics of a driven transmon with steady-state photon number $\bar{n}_{r}(t\xrightarrow{}\infty)=400$, $\kappa/2\pi=\SI{1}{MHz}$ (negative detuning), and $\kappa/2\pi=\SI{10}{MHz}$ (positive detuning); see \cref{sec: Floquet with just transmon} for details. Note the change in color scale between the top and bottom.
 }
 \label{fig:comparison_quantum_floquet}
\end{figure}

Using the fully quantum branch analysis, we now introduce critical photon numbers $n_{{\rm crit}, i_t}^q$ at which ionization of the transmon state $i_t$ is expected to occur. These critical numbers are defined as the minimum photon number $N_r$ at which the average transmon population reaches $N_t = 2$ for the ground-state branch and $N_t = 3$ for the excited-state branch. These choices capture large changes caused by resonances but avoid registering the slow and smooth increase in $N_t$ with photon number that is observed before ionization; see, e.g., the red lines in~\cref{fig:resonance_gate_charge}. Given the sensitivity of ionization to gate charge, the critical photon numbers we report are averaged over 100 realizations of $n_g$ uniformly distributed between 0 and 0.5.

\Cref{fig:comparison_quantum_floquet} shows the resulting $n_g$-averaged critical photon number $\bar n^q_{\mathrm{crit},i_t}$ as a function of the qubit-resonator detuning (pink line). The top panels correspond to the ground state and the bottom panels to the excited state. We exclude a range of detunings around $\Delta=0$ where the dispersive approximation does not hold (dashed region). The shaded pink area shows the distribution of $n_{{\rm crit}, 0_t}^q$ and $n_{{\rm crit}, 1_t}^q$ between the tenth and 90th percentiles for the realizations of $n_g$, emphasizing the strong dependence of ionization on gate charge. The dashed blue line is the usual Jaynes-Cummings critical photon number $n_{{\rm crit}, i_t}$ related to the breakdown of the dispersive approximation (see \cref{App. Sec.:dispersive_ncrit}). Apart from a few values of $n_g$, the onset of ionization is well above $n_{{\rm crit}, i_t}$. This observation is in agreement with Refs.~\cite{Jeffrey2014Fast,Sunada2022}, where a $\gtrsim 99 \%$ dispersive readout fidelity was obtained at an average photon number larger than $n_{\rm crit}$.

For comparison, the background shows the average transmon population obtained from numerical integration of Schrödinger's equation for a driven transmon, a quantity that is more numerically intensive to obtain than the critical photon numbers (pink line) but more closely resembling experiments. These time-dynamics simulations do not include a resonator mode. In \cref{sec: Floquet with just transmon}, we indeed show that, with the appropriate choice of time-dependent drive amplitude, the dynamics of a driven transmon captures the essential physics of the fully quantum model. Because of the strong dependence on $n_g$, the results are also averaged over 100 realizations of the gate charge. The agreement between the critical photon numbers extracted from the branch analysis and the sharp increase in the average population of the simulated driven transmon is excellent, confirming that the $n_{{\rm crit}, i_t}^q$ are useful and numerically simple to compute proxies for the onset of ionization.

The critical photon numbers $n_{{\rm crit}, i_t}^q$ exhibit a complex behavior as a function of detuning which can be understood from the phenomenology discussed in the previous sections. Focusing first on the case of negative detuning, we first observe the expected overall increase of $n_{{\rm crit}, i_t}^q$ with $|\Delta|$ due to the decrease of the $\chi_{i_t}$ shifts. As these shifts become smaller in magnitude, the slope of the modular energies decreases, pushing the energy collisions responsible for branch swapping to larger photon numbers and, consequently, increasing $n_{{\rm crit}, 0_t}^q$ and $n_{{\rm crit}, 1_t}^q$. On the other hand, the large detuning-dependent dips are due to branch swapping caused by avoided crossings between the ground-state branch [\cref{fig:comparison_quantum_floquet}(a)] or excited-state branch [\cref{fig:comparison_quantum_floquet}(c)] and other branches. Increasing the detuning has the effect of sweeping through these resonances, resulting in the observed dips. For example, the three large dips in \cref{fig:comparison_quantum_floquet}(a) arise from resonances in the modular spectrum between the ground-state branch and (from small to large negative detuning) branches $B_{7_t}$, $B_{6_t}$, and $B_{5_t}$; see \cref{App. Sec.: Non-perturbative corrections} for details. 

At positive detuning, we also observe the expected overall increase of the critical photon numbers with detuning. On top of this behavior are sharp dips resulting from multiphoton resonances between the ground state [\cref{fig:comparison_quantum_floquet}(b)] or excited state [\cref{fig:comparison_quantum_floquet}(d)] and a quasiresonant partner state in the inverted potential in \cref{fig:pos_branch_analysis_1Dlattice}(a). At larger detuning, the ground and excited states no longer have a quasiresonant partner; see \cref{fig:pos_branch_analysis_1Dlattice}(b). As a result, this mechanism precipitating ionization is no longer active, and the increase of the critical photon number with detuning becomes more monotonic. 
 
\section{Floquet analysis of the driven transmon} 
\label{sec: Floquet with just transmon}

During qubit readout, the drive $\varepsilon_d$ on the resonator displaces the resonator field to a coherent state with amplitude $\alpha(t)$ which, in turn, acts as a classical drive on the transmon~\cite{Lledo2023Cloaking, Munozarias2023Qubit}. By making a displacement transformation $\hat a \rightarrow \hat a + \alpha(t)$ on the master equation [\cref{eq:Full_ME}] and subsequently ignoring the quantum fluctuations of the resonator, we arrive at the simplified semiclassical picture of a driven transmon with Hamiltonian (see \cref{App. Sec.: Effective drive on the transmon}):
\begin{equation} \label{eq: driven Transmon Hamiltonian}
\begin{split}
\hat H(t) =& 4E_C(\hat n_t - n_g)^2 - E_J\cos(\hat \varphi_t) + \mathcal E_t(t) \hat n_t \\
=& \hat H_t + \mathcal E_t(t) \hat n_t.
\end{split}
\end{equation}
Assuming the resonator drive frequency $\omega_d$ to be close to the resonator frequency $\omega_r$, the classical drive on the transmon is approximately $\mathcal E_t(t) \approx \varepsilon_t(t)\cos(\omega_d t)$, with a time-dependent amplitude $\varepsilon_t(t) = 2g\sqrt{\bar n_r(t)}$, where 
\begin{equation}\label{eq:Floquet_nr}
\bar n_r(t) =\left(\frac{\varepsilon_{d}}{\kappa}\right)^2 (1-e^{-\kappa t/2})^2 
\end{equation}
is the average number of photons in the resonator; see \cref{App. Sec.: Effective drive on the transmon} for details. In this model, the resonator frequency $\omega_r$ of the fully quantum model is replaced by the drive frequency $\omega_d$. As a result, we place the qubit frequency below or above $\omega_d$ when considering negative or positive detuning, respectively.

The effective drive on the transmon originates from the displacement of the transmon-resonator coupling. Thus, its amplitude is proportional to the coupling strength $g$. With the usual large values of $g$ in circuit QED, this can lead to a very large effective drive amplitude on the transmon. For example, for the value $g/2\pi = \SI{120}{MHz}$ used in this paper, we have $\varepsilon_t/2\pi \sim \SI{750}{MHz}$ for $\bar n_r = 10$ photons in the resonator. This is significantly larger than the transmon anharmonicity $\alpha$. Even though it is off resonant, this effective drive cannot be treated perturbatively and can be expected to lead to leakage out of the computational subspace. In this section, we show that this simplified semiclassical model leads to the same predictions for transmon ionization as the fully quantum model. The essence of the success of this simple approach is that it preserves the full cosine potential of the transmon qubit. 

Because the effective drive amplitude $\varepsilon_t(t) \propto \sqrt{\bar n_r(t)}$ changes on a timescale $1/\kappa$ that is much larger than the period of the drive ($T = 2\pi/\omega_d$), our analysis relies on the instantaneous Floquet spectrum~\cite{Breuer1989QuantumPhases}. The instantaneous spectrum is obtained by taking $\varepsilon_t$ to be constant, such that $\hat H(t)$ is periodic. The solution to Schrödinger’s equation can then be expressed as $|\psi(t)\rangle 
= \sum_{i_t} c_{i_t} e^{-i\epsilon_{i_t} t} |\phi_{i_t}(t)\rangle$, where $\epsilon_{i_t}$ is a Floquet quasienergy and $|\phi_{i_t}(t)\rangle = |\phi_{i_t}(t+T)\rangle$ is a Floquet mode~\cite{Grifoni1998}. The phases $\exp(-i\epsilon_{i_t}T)$ and the Floquet modes $\ket{\phi_{i_t}(t)}$ are the eigenvalues and eigenvectors of the propagator over one period of the drive, $\hat U(t+T, t)|\phi_{i_t}(t)\rangle = e^{-i\epsilon_{i_t} T} |\phi_{i_t}(t)\rangle$. These eigenvalues can also be obtained from a static Hamiltonian whose form is reminiscent of the transmon-resonator Hamiltonian; see \cref{App. Sec.: Floquet formalism}. Importantly, as the phases $\exp(-i\epsilon_{i_t}T)$ remain invariant under a shift of $\epsilon_{i_t}$ by an integer multiple of $\omega_d$, the quasienergies $\epsilon_{i_t}$ are defined only modulo $\omega_d$. Thus, quasienergies are analogous to quasimomenta in Bloch theory, and they are folded in a first ``Brillouin zone'' $-\omega_d/2 \le\epsilon_{i_t} \le \omega_d/2$. Importantly, the Floquet modes and quasienergies are functions of the drive amplitude $\varepsilon_t$.

Just as in the previous section, we sort the Floquet quasienergies $\epsilon_{i_t}[\varepsilon_t]$ and modes $|\phi_{i_t}[\varepsilon_t]\rangle$ into transmon branches. Our labeling procedure at finite drive is performed by smoothly increasing the drive amplitude to connect the eigenstates of the propagator to those at zero drive. This is made possible by the fact that, at zero drive, Floquet modes are simply the bare transmon eigenstates, $\ket{\phi_{i_t}[\varepsilon_t=0]}=\ket{i_t}$, and the quasienergies are the bare transmon energies modulo the drive frequency, $\epsilon_{i_t}[\varepsilon_t=0] = E_{i_t} \mod \omega_d$. Thus, the Floquet quasienergy spectrum is akin to the modular energy spectrum introduced above for the fully quantum model. Both spectra capture changes in (quasi)energy with photon number. Moreover, thanks to the folding of the spectrum present in both cases, there can be avoided crossings between states corresponding to low- and high-energy transmon states; see e.g.~\cref{fig:Floquet negative detuning}(b). Thus, avoided crossings in the quasienergy spectrum are linked to resonances between ac-Stark-shifted transmon states, up to $k$ drive photons of energy $\omega_d$~\cite{Xiao2023Diagrammatic}. More details on the correspondence with the fully quantum approach are given in \cref{App. Sec.: Floquet formalism}.

For unbounded Hamiltonians such as the transmon Hamiltonian, however, there are avoided crossings with quasienergy gaps of arbitrarily small sizes~\cite{Hone1997Time-dependent}. To prevent capturing very small avoided crossings that are not relevant to transmon ionization, we numerically obtain the Floquet spectrum as a function of $\varepsilon_t$ by choosing a finite increment $\delta \varepsilon_t$ for tracking the Floquet branches $B_{i_t}$. This freedom in the choice of $\delta\varepsilon_t$ contrasts with the branch construction of the fully quantum model where one resonator photon is added at each step by default, corresponding to an increment $\sim g/\sqrt{n_r}$. Here, we take $\delta \varepsilon_t/2\pi = \SI{10}{MHz}$, which captures the avoided crossing seen in the modular spectrum of the fully quantum model. Another reason for this choice is that the effective drive amplitude $\varepsilon_t(t)$ is linked to the photon population of the resonator, which in a qubit readout goes from 0 to a finite population $n_r(t)$ on a timescale set by $\kappa$. As such, the size of the increment $\delta \varepsilon_t$ can, to some extent, be linked to the system's dynamics. Thus, we have taken $\delta\varepsilon_t$ to be close to the value of $\kappa/2\pi = \SI{7.95}{MHz}$ used in our time-dependent simulations.

\subsection{Negative detuning \texorpdfstring{$\omega_q < \omega_r$}
 {\unichar{"03C9}\unichar{"1D45E} < \unichar{"03C9}\unichar{"1D45F}}}
\label{subsec: floquet negative detuning}
\begin{figure}[t]
 \centering
 \includegraphics[width=\linewidth]{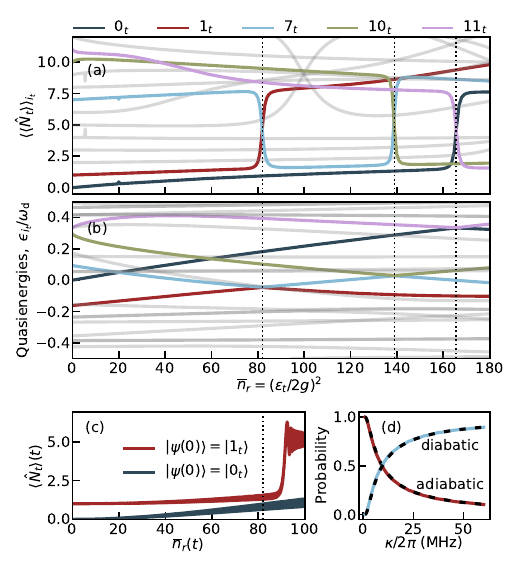}
 \caption{{\bf Negative detuning.} (a) Period-averaged transmon population of the Floquet modes and (b) quasienergy spectrum of the Floquet modes as a function of the average resonator photon number $\bar n_r = (\varepsilon_t/2g)^2$ for a negative detuning $\Delta/2\pi = -\SI{1.217}{GHz}$. In (a,b), branches $B_{0_t}$, $B_{1_t}$, $B_{7_t}$, $B_{10_t}$, and $B_{11_t}$ are highlighted in color (see the legend), and the others are gray. In (b), energies are measured from the qubit ground state energy and then folded into the interval $[-\omega_d/2, \omega_d/2]$. The transmon parameters are the same as in \cref{fig: fully quantum negative detuning}. Swapping of the branch populations occurs at avoided crossings in the quasienergy spectrum (dotted vertical lines). The transmon parameters are the same as in \cref{fig: fully quantum negative detuning}. (c) Time evolution of the average transmon population under a drive of amplitude $\varepsilon_t(t) = 2g \sqrt{100} (1-e^{-\kappa t/2})$, with $\kappa/2\pi = \SI{7.95}{MHz}$ and $\omega_d/2\pi = \SI{7.515}{GHz}$. The transmon is initialized in $\ket{0_t}$ (blue line) or $\ket{1_t}$ (red line).
 (d) Probability $|\bra{\phi_{i_t}[\varepsilon_t(t_f)]} \hat{\mathcal U}(t_f)\ket{1_t}|^2$ of being in the Floquet mode $i_t=1_t$ (red line, adiabatic) or $i_t=7_t$ (light blue line, diabatic) versus $\kappa$ at the final time $t_f\approx 10/\kappa$ of the evolution. The dashed black lines are the Landau-Zener predictions; see \cref{App. subsec: Floquet-Landau-Zener transition probability}.
 } 
 \label{fig:Floquet negative detuning}
\end{figure}

We saw in \cref{subsec: fully quantum negative detuning} that the opposite signs of the ac-Stark shifts of pairs of transmon branches (one deep in the cosine potential well and the other near the top of the well) result in avoided crossings in the modular energies. These avoided crossings -- which can be precipitated by the early hybridization of states at the top of the well -- lead to transmon ionization during readout. We now show that the same phenomenology is present in the simplified semiclassical model of a driven transmon [\cref{eq: driven Transmon Hamiltonian}], with its full cosine potential.

\Cref{fig:Floquet negative detuning}(b) shows the numerically obtained quasienergy spectrum as a function of the transmon drive amplitude, here expressed in terms of the equivalent average resonator photon number $\bar n_r = (\varepsilon_t/2g)^2$. The similarity with the modular energy spectrum in \cref{fig: fully quantum negative detuning}(b) obtained from the fully quantum model is remarkable. In correspondence with \cref{fig: fully quantum negative detuning}(b), the quasienergies of states at the bottom and near the top of the cosine potential well move in opposite directions with $\bar n_r$ (see \cref{App. Sec.: Floquet formalism}), leading to avoided crossings at finite drive amplitudes. As an example, the quasienergies of branches $B_{1_t}$ (red line) and $B_{7_t}$ (light blue line) approach each other until an avoided crossing is formed at $\bar n_r\approx 82$ photons. 

To illustrate the abrupt change in the character of the Floquet modes at the avoided crossings, we show in \cref{fig:Floquet negative detuning}(a) the period-averaged transmon population
\begin{equation} \label{eq: period-averaged transmon excitation in each Floquet mode}
\langle \langle \hat N_t \rangle \rangle_{i_t} = \frac{1}{T}\int_0^T d\tau\, \sum_{j_t} j_t |\langle j_t|\phi_{i_t}(\tau)\rangle|^2,
\end{equation} 
for each Floquet branch as a function of $\bar n_r$. The similarity with the transmon population of the branches computed for the full transmon-resonator system of \cref{fig: fully quantum negative detuning}(a) is again remarkable, with branch swapping occurring at the avoided crossings identified in \cref{fig:Floquet negative detuning}(b). Note also the presence of a small feature at $\bar n_r \approx 22$ of the period-averaged transmon population corresponding to the branches $B_{0_t}$ (dark blue line) and $B_{7_t}$ (light blue line). This feature is due to a very small avoided crossing of the same two branches also at $\bar n_r \approx 22$, which, with our choice of increment $\delta \varepsilon_t$, is not resolved and is shown as a crossing; see \cref{fig:Floquet negative detuning}(b). We confirm below that, with our choice of $\kappa$, this resonance does not play a role in the dynamics of the system.

To confirm that the observed resonances are responsible for transmon ionization in this semiclassical model, we now turn to time dynamics obtained under the Hamiltonian of \cref{eq: driven Transmon Hamiltonian} together with resonator decay $\kappa$; see \cref{fig:Floquet negative detuning}(c). Here, $\varepsilon_t$ is no longer a static control parameter but changes in time following $\varepsilon_t(t) = 2g\sqrt{\bar n_r(t)}$, with the average photon number given by \cref{eq:Floquet_nr}. After preparing the transmon in the bare state $\ket{\psi(0)} = \ket{i_t}$, the state adiabatically follows the instantaneous Floquet mode of the $i_t$th branch, $\ket{\phi_{i_t}[\varepsilon_t(t)](t)}$, as the drive amplitude is increased and until an avoided crossing with branch $B_{j_t}$ is met. Following the usual Landau-Zener argument~\cite{Grifoni1998}, at that point the state diabatically switches branch or adiabatically remains in the branch. This transition occurs with the Landau-Zener probability $P_\mathrm{LZ}=\exp(-\pi \Delta_{\rm ac}^2/2v)$, where $\Delta_{\rm ac}$ is the quasienergy gap at the avoided crossing and $v(\varepsilon_t)$ is the speed of passage through the resonance~\cite{Drese1999Floquet, Oliver2005MachZender,Shevchenko2010LandauZenerStuckelberg, Ikeda2022Floquet-Landau-Zener}; see \cref{App. subsec: Floquet-Landau-Zener transition probability}. Importantly, and as explained in \cref{sec:Fully quantum model}, ionization is averted if the state transitions diabatically through the crossing. By contrast, transitioning adiabatically means that the transmon ionizes. 

In \cref{fig:Floquet negative detuning}(c), we show a parametric plot of the transmon population as a function of time, here expressed in terms of $\bar n_r(t)$. When initially prepared in $\ket{0_t}$ (blue line), the state of the transmon follows branch $B_{0_t}$, going straight through the weak avoided crossing at $\bar n_r \sim 22$ that our finite-step-size Floquet tracking purposely did not capture. In this case, the drive does not induce transitions in the transmon. On the other hand, when initially prepared in $\ket{1_t}$ (red line), the state follows mostly adiabatically branch $B_{1_t}$ through the avoided crossing at $\bar n_r\approx 82$, leading to a transition to high-energy transmon states. \Cref{fig:Floquet negative detuning}(d) shows the probability of transitioning adiabatically or diabatically through the avoided crossing as a function of the rate at which $\varepsilon_t(t)$ varies. More precisely, we show as a function of $\kappa$ [which controls $\varepsilon_t(t)$; see \cref{eq:Floquet_nr}] the probability of the state being in branch $B_{i_t}=B_{1_t}$ (red line) or in branch $B_{i_t}=B_{7_t}$ (light blue line) at the final time $t_f \approx 10/\kappa$ of the evolution. By increasing the value of $\kappa$ and, thus, the speed at which resonances are crossed, the Landau-Zener probability for a diabatic crossing can be tuned from zero to almost one~\cite{Shillito2022Dynamics}. On top of these curves, we show the transition probability obtained from the Floquet-Landau-Zener formula (dashed black line) in perfect agreement with the time-dynamics simulations; see \cref{App. subsec: Floquet-Landau-Zener transition probability}. 

The onset of ionization, therefore, depends on the effective speed at which resonances are crossed. According to \cref{eq:Floquet_nr}, the drive ramp-up is faster at low photon numbers and slows down when approaching the steady-state photon number. Thus, we expect resonances that occur at the steady-state photon number to be more easily resolved. Consequently, ionization can be pushed back by diabatically crossing the resonance in the transient. This is observed experimentally in Ref.~\cite{Sank2016Measurement}, wherein the probability of qubit ionization exhibits an initial rise at a certain drive power, followed by a subsequent decline at higher powers. However, it is important to note that ionization can also happen at an avoided crossing when the resonator field ramps down after the measurement, and this must also be accounted for to explain experimental observations.

\subsection{Positive detuning \texorpdfstring{$\omega_q > \omega_r$}
 {\unichar{"03C9}\unichar{"1D45E} > \unichar{"03C9}\unichar{"1D45F}}}
\label{subsec: floquet positive detuning}

As discussed in \cref{subsec: fully quantum positive detuning}, because of the negative anharmonicity of the transmon, the branch swapping observed at negative detuning is replaced by branch bunching at positive detuning. We now show that this phenomenology is captured by the simplified semiclassical model of a driven transmon. Because ionization typically occurs at smaller resonator photon numbers at positive detuning than at negative detuning, larger discrepancies between the fully quantum model and the semiclassical model are expected.

\begin{figure}[t!]
 \centering
 \includegraphics[width=\linewidth]{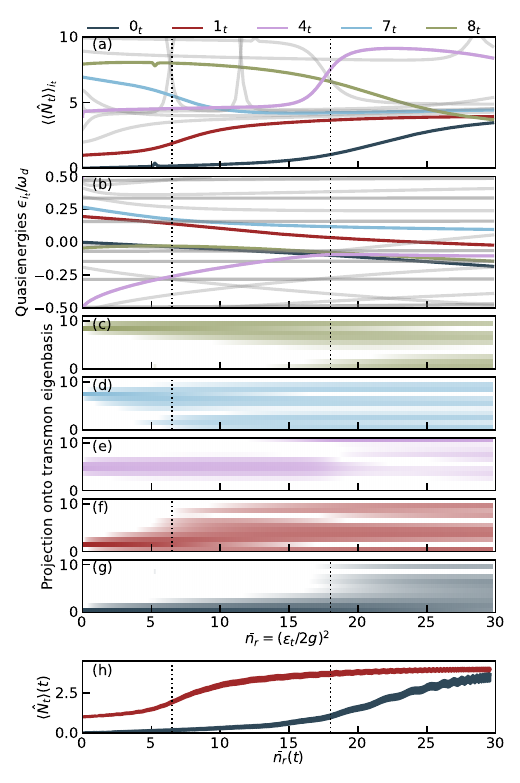}
 \caption{
 {\bf Positive detuning.} 
 (a) Period-averaged transmon population of the Floquet modes, (b) quasienergy spectrum of the Floquet modes, and (c-g) projections of the Floquet modes on the bare transmon basis as a function of the average resonator photon number $\bar n_r = (\varepsilon_t/2g)^2$ for a positive detuning $\Delta/2\pi =\SI{1.031}{GHz}$. In (a,b), branches $B_{0_t}$, $B_{1_t}$, $B_{4_t}$, $B_{7_t}$, and $B_{8_t}$ are highlighted in color (see the legend), and the others are gray. In (b), energies are measured from the qubit ground state energy and then folded into the interval $[-\omega_d/2, \omega_d/2]$. In (c-g), the projections of $8_t$, $7_t$, $4_t$, $1_t$, and $0_t$ are plotted on a logarithmic color bar scale with the same color coding. The dotted vertical lines indicate the average resonator photon number at which the population bunching in $B_{0_t}$ and $B_{1_t}$ dramatically increases. The transmon parameters are the same as in \cref{fig:pos_branch_analysis}. 
 (h) Time evolution of the average transmon population under a drive of amplitude $\varepsilon_t(t) = 2g \sqrt{30} (1-e^{-\kappa t/2})$, with $\kappa/2\pi = \SI{11.98}{MHz}$ (set to match the full $\chi$ shift at zero photons) and $\omega_d/2\pi = \SI{5.267}{GHz}$. The transmon is initialized in $\ket{0_t}$ (blue line) or $\ket{1_t}$ (red line).
}
\label{fig:Floquet positive detuning}
\end{figure}

\Cref{fig:Floquet positive detuning}(a,b) show the period-averaged transmon population of the Floquet branches and the quasienergy spectrum, respectively, versus $\bar n_r$ and obtained for the same parameters as in \cref{subsec: fully quantum positive detuning}. Although there are some discrepancies, especially at small photon numbers, the agreement with the modular spectrum of the fully quantum model is nevertheless excellent; see \cref{fig:pos_branch_analysis}. As in the fully quantum model, rather than branch swapping we observe in \cref{fig:Floquet positive detuning}(a) the expected branch bunching at $\langle\langle \hat N_t \rangle\rangle \sim 4$. The branch bunching also correctly correlates with avoided crossings in the quasienergy spectrum. For example, at $\bar n_r \approx 6.4$, the quasienergies of branches $B_{1_t}$ and $B_{7_t}$ form a strong and wide avoided crossing, at which point we also observe the bunching of these two branches. Interestingly, a weak avoided crossing of the quasienergies associated with $0_t$ and $8_t$ at $\bar n_r\approx 5.3$ is not captured with our finite tracking increment of $\delta\varepsilon_r/2\pi = \SI{10}{MHz}$. While the $1_t$-$7_t$ transition requires the absorption of six drive photons, the $0_t$-$8_t$ transition requires eight. Hence, it is more strongly suppressed and is not observed at this low drive amplitude. This higher-order transition, however, becomes relevant at larger drive amplitudes. Indeed, at $\bar n_r \approx 18$ a weak level repulsion of the quasienergies associated with $0_t$ and $8_t$ is observed which correlates with the start of the bunching of these two states. 

The correlation between branch bunching and avoided crossings is further confirmed by examining the projection of the Floquet modes onto the bare transmon basis; see \cref{fig:Floquet positive detuning}(c,d,f,g). There, we see that the Floquet modes associated with $0_t$ and $8_t$ show a strong delocalization in the bare transmon basis at $\bar n_r \approx 18$. Similarly, for $1_t$ and $7_t$ this occurs at $\bar n_r \approx 6.5$, matching the presence of the avoided crossing in the quasienergy spectrum. To quantify the delocalization of the modes swallowed into the bunching layer, we compute the inverse participation ratio $\sum_{j_t} |\langle j_t | \phi_{i_t}\rangle|^4$ \cite{Kramer1993}, which here measures the delocalization of a Floquet mode $\ket{\phi_{i_t}}$ on the transmon basis. 
At $\bar n_r\approx 11$, the inverse participation ratio is approximately $0.14 \approx 1/7$ for both $\ket{\phi_{7_t}}$ and $\ket{\phi_{1_t}}$, corresponding to a delocalization of these modes across at least seven bare transmon states.

To confirm that the presence of avoided crossings in the quasienergy spectrum is an accurate predictor of the system's dynamics, in \cref{fig:Floquet positive detuning}(h) we show the time-dependent transmon population as a function of time, here parametrized by the instantaneous average photon number $\bar n_r(t)$. When the transmon is initialized in $\ket{0_t}$ (blue line), the state follows branch $B_{0_t}$, avoiding an early measurement-induced ionization at $\bar n_r\approx 5.3$.
As a result, no feature is observed in $\langle \hat N_t\rangle(t)$ at that average photon number, except for a slow overall increase of the transmon population due to the gradual hybridization of branch $B_{0_t}$ with other branches. At $\bar n_r\approx 18$, the transmon population increases, as expected from the delocalization of that branch; see \cref{fig:Floquet positive detuning}(g). On the other hand, when initialized in $\ket{1_t}$ (red line) the system mostly adiabatically transitions at the large avoided crossing at $\bar n_r\approx 6.4$, thereby following branch $B_{1_t}$ with a high probability. This results in measurement-induced transitions and in the observed rapid increase of $\langle \hat N_t\rangle(t)$.

\subsection{Comparison to the fully quantum model}
\label{subsec:comparison_quantum_floquet}

We define the critical photon numbers in the Floquet analysis similarly as in the branch analysis of the fully quantum model. Thus, the Floquet critical photon numbers for the ground state $n_{{\rm crit}, 0_t}^f$ and for the excited state $n_{{\rm crit}, 1_t}^f$ are defined as the smallest value of $\bar n_r$ for which $\langle\langle \hat N_{t}\rangle\rangle_{0_t}= 2$ and $\langle\langle \hat N_{t}\rangle\rangle_{1_t}=3$, respectively. In \cref{fig:comparison_quantum_floquet}, we compare the extracted $n_{{\rm crit}, i_t}^f$ (red line) with $n_{{\rm crit}, i_t}^q$ of the fully quantum model (pink line) for a range of detunings. The agreement between the two methods is excellent for both the average values and the fluctuations with charge noise (see the red and pink shaded areas). This agreement confirms the validity of treating the driven resonator as an effective classical field driving the transmon. Numerically, this approach is advantageous because of its simplicity and efficiency.

As previously mentioned, in \cref{fig:comparison_quantum_floquet} we also compare the critical photon numbers with the dynamics of the driven transmon (colored background). Because avoided crossings are very narrow in the negative detuning regime, we take a small $\kappa/2\pi=\SI{1}{MHz}$ to ensure a slow ramp-up of the average photon number in the resonator. In this way, the system slowly crosses the resonances responsible for $\bar n_{{\rm crit}, i_t}^q$ and $\bar n_{{\rm crit}, i_t}^f$, leading to ionization in the time dynamics. At positive detuning, resonances are wider and a faster ramp-up with $\kappa/2\pi=\SI{10}{MHz}$ still results in ionization around $\bar n_{{\rm crit}, i_t}^q$ and $\bar n_{{\rm crit}, i_t}^f$. When comparing to experiments, the values of $\kappa$ and $\varepsilon_d$ used in the time-dynamical simulations can be adjusted to match the speed at which the resonator field builds up. For large speeds, early weak resonances are crossed rapidly and, therefore, do not lead to ionization. The step size $\delta\varepsilon_t$ in the Floquet tracking should be adjusted accordingly.

\section{Nonlinear dynamics of a driven classical transmon}
\label{sec:chaos}

We showed in \cref{sec: Floquet with just transmon} that a simplified semiclassical model of a transmon driven by a classical monochromatic drive is enough to understand and predict the onset of ionization. In this section, we go one step further and investigate a model where the transmon itself is treated as a classical object. As in the previous sections, we account for the full nonlinearity of the transmon cosine potential and treat the readout drive nonperturbatively. This fully classical description, combined with the Bohr-Sommerfeld quantization rule, was shown in Ref.~\cite{Cohen2023Reminiscence} to capture the main features of ionizing transitions. While this work points out the relevance of these transitions for readout, a quantitative analysis as well as a comparison to other methods remains lacking. Moreover, the qualitative difference between negative and positive transmon-resonator detunings was not fully explored, since the Bohr-Sommerfeld quantization used there applies only to positive detunings where a single large resonance dominates (see below). Here, we further clarify the classical mechanisms of ionization for both negative and positive detuning, and we explicitly connect them to the quantum phenomenology discussed in \cref{sec:Fully quantum model} and \cref{sec: Floquet with just transmon}.

\subsection{General properties of the classical model}

\begin{figure*}[t]
 \centering
 \includegraphics[width=\textwidth]{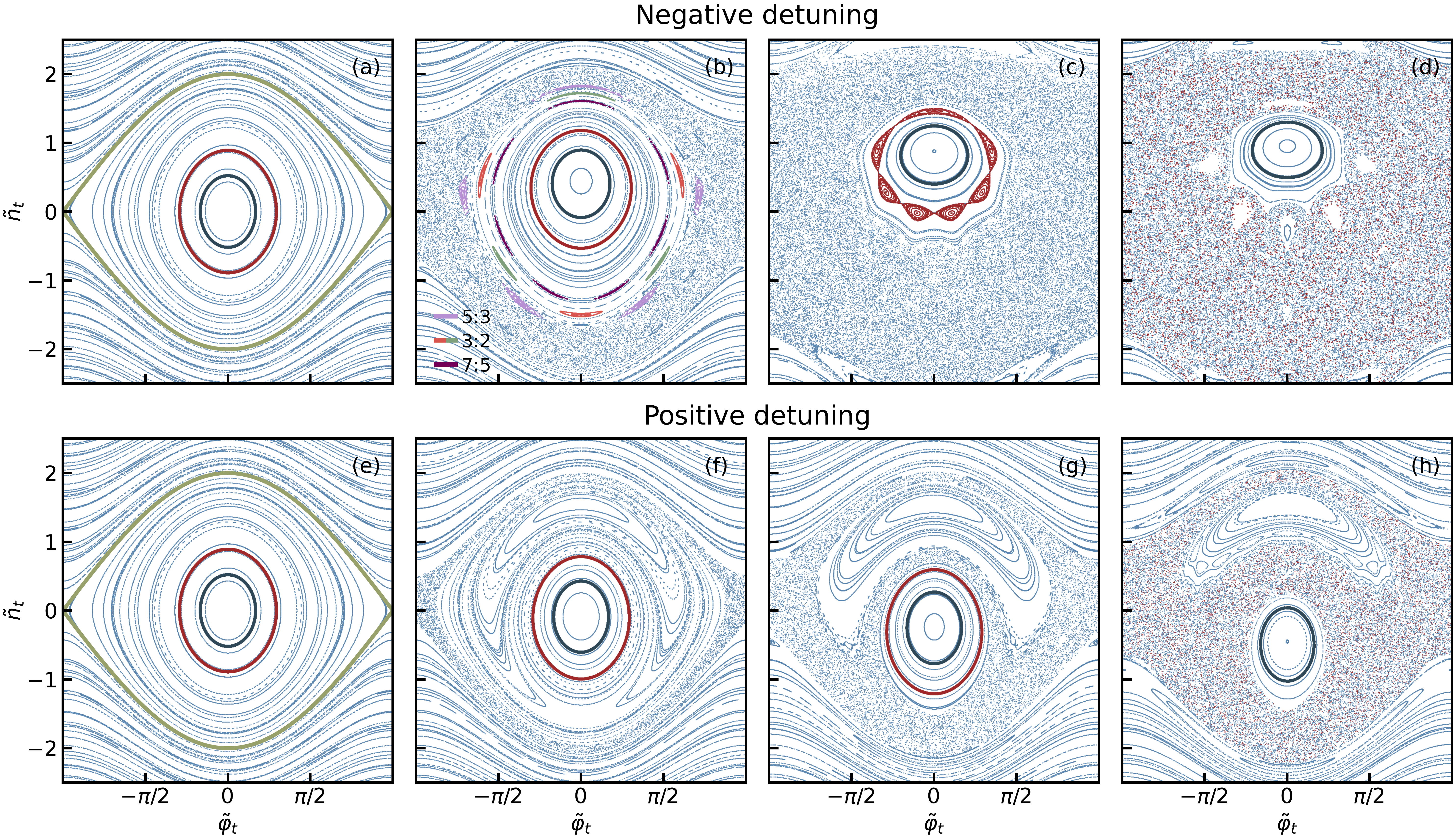}
 \caption{
 Poincaré sections for different values of the drive amplitude $\Tilde{\varepsilon}_t$ at negative (top row) and positive (bottom row) detuning. For negative detuning (a-d), the drive frequency is $\Tilde{\omega}_d = 1.152$ ($\omega_d/2\pi = \SI{7.515}{GHz}$, $\Delta/2\pi = -\SI{1.202}{GHz}$) and the drive amplitudes are, from left to right, $\Tilde{\varepsilon}_t \in \{0, 0.167, 0.501, 0.624\}$ ($\bar{n}_r \in \{0, 20.6, 185.6, 287.6\}$). For positive detuning (e-h), the drive frequency is $\Tilde{\omega}_d = 0.807$ ($\omega_d/2\pi = \SI{5.267}{GHz}$, $\Delta/2\pi = \SI{0.998}{GHz}$) and the drive amplitudes are, from left to right, $\Tilde{\varepsilon}_t \in \{0, 0.029, 0.081, 0.152\}$ ($\bar{n}_r \in \{0, 0.63, 4.88, 17.02\}$). The Bohr-Sommerfeld orbits of the qubit ground and excited states are shown for $E_J/E_C = 110$ in dark blue and dark red, respectively. At zero drive amplitude (a,e), the separatrix is shown in olive. In (b), $\nm{5}{3}$ and $\nm{7}{5}$ resonances are highlighted in purple and plum red, respectively. A pair of $\nm{3}{2}$ resonances are also highlighted in orange and pale green. In (c), the $\nm{7}{5}$ resonance collides with the Bohr-Sommerfeld orbit of the excited state. This is illustrated by coloring the resonance in red. In (d), the merging of the excited state with the chaotic layer is assisted by a pair of $\nm{3}{2}$ resonances that reduce the regular area available to the excited states. In (f-h), a $\nm{1}{1}$ resonance occupies a large fraction of the region inside the main separatrix. At large drive amplitudes (d,h), the excited state is engulfed by the chaotic layer. This is illustrated by the red points in the chaotic layer.
 }
 \label{fig:poincare_maps_2x4} 
\end{figure*}

The full nonlinear dynamics of the classical transmon is governed by a dimensionless Hamiltonian that is formally identical to that of a driven classical pendulum~\cite{Graham1991Level,Koch2007ChargeInsensitive}:
\begin{equation} \label{eq: classical Hamiltonian}
\begin{split}
 \Tilde{H}( \Tilde{t} ) &= \frac{1}{2} \Tilde{n}_t^2 - \cos{\Tilde{\varphi}_t} + \Tilde{\varepsilon}_t \cos({\Tilde{\omega}_d \Tilde{t}})\Tilde{n}_t \\
       &= \Tilde{H}_t + \Tilde{\varepsilon}_t \cos({\Tilde{\omega}_d \Tilde{t}})\Tilde{n}_t.
\end{split}
\end{equation}
Here, $\Tilde{H}_{t} = H_{t}/E_J$ and $\Tilde{t} = \omega_p t$ are the rescaled Hamiltonian and time, respectively, with $\omega_p = \sqrt{8E_J E_C}$ the transmon plasma frequency. Moreover, $\Tilde{\varepsilon}_t = \varepsilon_t/\omega_p$ and $\Tilde{\omega}_d = \omega_d/\omega_p$ are the rescaled transmon drive amplitude and drive frequency, respectively. The rescaled phase-space coordinates are $\Tilde{\varphi}_t = \varphi_t$ and $\Tilde{n}_t = z n_t$ with Poisson bracket $\{\Tilde{\varphi}_t ,\Tilde{n}_t \} = z$, where $z= \sqrt{8E_C/E_J}$ is the transmon impedance~\cite{Cohen2023Reminiscence}. We omitted the gate charge $n_g$ from \cref{eq: classical Hamiltonian}, since a static gate charge does not affect classical dynamics. Moreover, as in \cref{sec: Floquet with just transmon}, the drive frequency $\omega_d$ is a surrogate for the resonator frequency of the fully quantum model.

Given the periodic time dependence of \cref{eq: classical Hamiltonian}, the solutions to Hamilton's equations are best studied using the Poincaré section, obtained by plotting the phase-space coordinates stroboscopically at multiples of the drive period $\Tilde{T} = 2\pi/\Tilde{\omega}_d$. In \cref{fig:poincare_maps_2x4}, we show the Poincaré sections for a negative detuning (top) and for a positive detuning (bottom) of the qubit frequency to the drive frequency. In each case, we show the Poincaré sections for increasing values of the drive amplitude $\Tilde{\varepsilon}_t$. These results are obtained for the same values of the transmon parameters as above, here corresponding to $\omega_p/2\pi = \SI{6.526}{GHz}$ and $z \sim 0.27$.

When $\Tilde{\varepsilon}_t = 0$, two distinct types of regular orbits fill phase space; see \cref{fig:poincare_maps_2x4}(a,e). For $\Tilde{H}_t < 1$, the system undergoes bounded periodic phase oscillations that are the classical analogs of the transmon bound states. In the Poincaré section, these bounded oscillations correspond to the near-circular orbits located in the center of phase space. Below, we refer to this set of orbits as the main regular region. On the other hand, for $\Tilde{H}_t > 1$ the system undergoes full $\pm2\pi$ phase rotations, corresponding to phase slips in the quantum model~\cite{Koch2007ChargeInsensitive}. This type of motion appears in the Poincaré section as nearly horizontal features that do not cross $\tilde n_t = 0$. Separating these two types of motion is the contour $\Tilde{H}_t = 1$ known as the separatrix [olive line in \cref{fig:poincare_maps_2x4}(a,e)]. In the vicinity of the separatrix, small changes to the initial conditions can lead to large changes in the system's dynamics.

In addition to bounded and unbounded oscillations, for $\Tilde{\varepsilon}_t > 0$ new types of motion are visible in the Poincaré sections. The first type of motion common to all panels in \cref{fig:poincare_maps_2x4} for which $\Tilde{\varepsilon}_t > 0$ is chaotic motion. Chaos emerges in the vicinity of the separatrix for an arbitrarily weak drive amplitude and results from the instability of the system with respect to small perturbations at energies close to $\tilde H_t = 1$. % the separatrix. 
The size of the phase-space region covered by chaotic motion generally increases with the drive amplitude~\cite{Zaslavsky2005, Wimberger2014}. Critically, this process reduces the area of phase space supporting regular bounded oscillatory motion. This reduction of the phase-space area available to the main regular region due to the emergence of chaotic layers was identified as a key phenomenon responsible for the onset of ionization in the classical model~\cite{Cohen2023Reminiscence}. We note that the emergence of chaos in the quantum model of \cref{eq: driven Transmon Hamiltonian} and the classical model of \cref{eq: classical Hamiltonian} were studied in Ref.~\cite{Graham1991Level}. 

The second type of motion seen in all panels in \cref{fig:poincare_maps_2x4} for which $\Tilde{\varepsilon}_t > 0$ is the presence of sets of islands surrounding the main regular region; see, e.g., the features highlighted in plum red, olive, and purple in \cref{fig:poincare_maps_2x4}(b), as well as the large croissant-shaped feature in \cref{fig:poincare_maps_2x4}(f-h). These features are subharmonic periodic orbits corresponding to $\nm{n}{m}$-nonlinear resonances occurring when $n \tilde \omega(\tilde I) = m \tilde \omega_d $~\cite{Zaslavsky2005}. In this expression, $\tilde \omega(\tilde I) = \partial \tilde H_t /\partial \tilde I$, with $\tilde I$ the rescaled action variable, is the energy-dependent frequency of periodic motion of the pendulum, which reduces to 1 at low energy. A $\nm{n}{m}$ resonance can be identified in phase space by a chain of $n$ islands. We show below that, in addition to the emergence of chaos, the presence of these resonances is a second key phenomenon responsible for the onset of ionization within the classical model. Importantly, the phase-space area occupied by a resonance depends on $n$, with increasing values of $n$ leading to smaller resonances, i.e., weaker resonances. The area occupied by a resonance and its distance from the center of the main regular region also generally depends on $\Tilde{\varepsilon}_t$. As a result, although there is typically an abundance of $\nm{n}{m}$ resonances due to the nonlinearity of $\tilde \omega(\tilde I)$~\cite{Zaslavsky2005}, not all of these play an equally important role in ionization.

\subsection{Bohr-Sommerfeld critical photon numbers}
\label{subsec:bohr_sommerfeld_critical_photon_numbers}

\begin{figure*}[t]
 \centering
 \includegraphics[width=\textwidth]{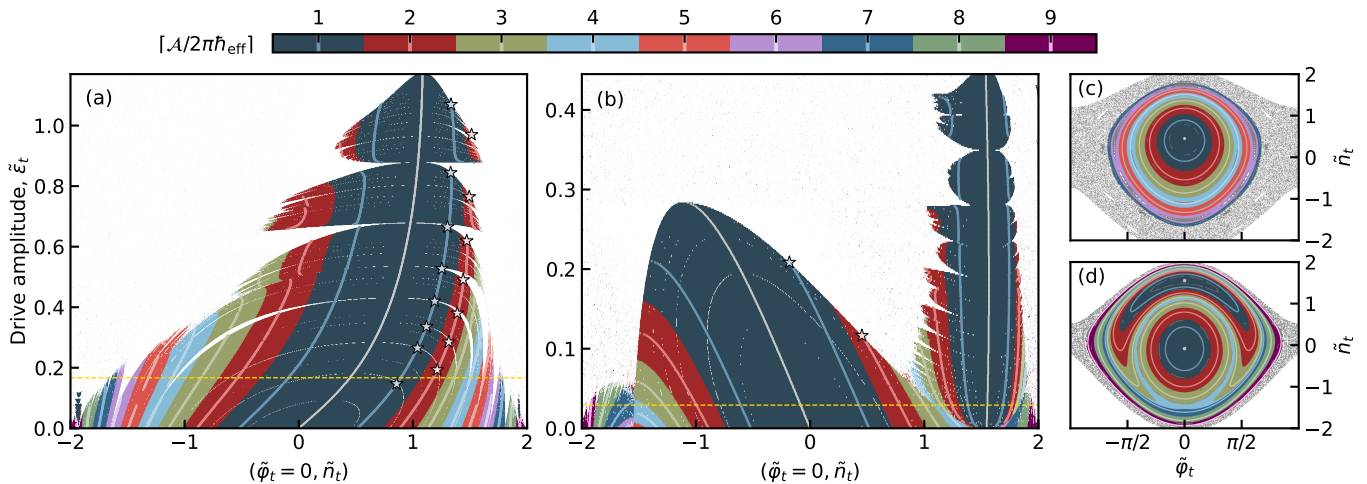}
 \caption{
 Cut $(\Tilde{\varphi}_t = 0, \Tilde{n}_t)$ of phase space as a function of $\Tilde{\varepsilon}_t$ for (a) negative detuning ($\Tilde{\omega}_d = 1.152$, $\omega_d/2\pi = \SI{7.515}{GHz}$, $\Delta/2\pi = -\SI{1.202}{GHz}$) and (b) positive detuning ($\Tilde{\omega}_d = 0.807$, $\omega_d/2\pi = \SI{5.267}{GHz}$, $\Delta/2\pi = \SI{0.998}{GHz}$). The horizontal axes scan the initial conditions $(\Tilde{\varphi}_t = 0, \Tilde{n}_t^0)$ with $\Tilde{n}_t^0 \in [-2.0, 2.0]$ varied in increments of $1/300$. The vertical axes scan the drive amplitudes in increments of $\delta\Tilde{\varepsilon}_t = 0.0015323$, corresponding to an increment of $\delta \varepsilon_t/2\pi = \SI{10}{MHz}$. The Bohr-Sommerfeld orbits are represented by pale colored lines, while dark regions with matching colors represent quantum fluctuations of $\pi\hbar_{\rm{eff}}$ on each side of the Bohr-Sommerfeld orbits. For $E_J/E_C=110$, the area occupied by a Bohr-Sommerfeld state is $2\pi\hbar_{\rm eff} = 1.69445$. The gray line centered at $\tilde{n}_t = 0$ at zero drive indicates the center of the main regular region, while the light blue and light red lines closest to it are the Bohr-Sommerfeld orbits associated with the qubit ground and excited states, respectively. In (b), the gray line at $\tilde{n}_t \approx 1.5$ indicates the center of the $\nm{1}{1}$ resonance. White regions indicate points where the algorithm in \cref{App. Sec.: Area algorithm chaos} fails to find a closed connected orbit due to chaotic or resonant motion. The points where the Bohr-Sommerfeld orbits associated with the qubit states stop existing (stars) yield the Bohr-Sommerfeld critical photon numbers $\{n_{\mathrm{crit},i_t}^{c}\}$. As discussed in the text, weak resonances are ignored. (c) and (d) show, using the same color code, the Poincaré sections corresponding to the drive amplitudes $\Tilde{\varepsilon}_t = 0.167$ ($\bar{n}_r = 20.6$) and $\Tilde{\varepsilon}_t = 0.029$ ($\bar{n}_r = 0.63$) indicated by the dashed yellow lines in (a) and (b), respectively. The other parameters in (c) and (d) are the same as in \cref{fig:poincare_maps_2x4}(b) and \cref{fig:poincare_maps_2x4}(f), respectively.}
 \label{fig:BohrSommerfeld_drive_heatmaps}
\end{figure*}

We now exploit the previous observations to predict the onset of ionization within the classical model. Interestingly, the relevance of chaotic behavior to the ionization of highly excited Rydberg atoms was already pointed out~\cite{Delone1983Highly, Casati1987Relevance, Wang1989Dynamics}. Here, we follow Ref.~\cite{Cohen2023Reminiscence} and use Bohr-Sommerfeld quantization to establish a correspondence between the stroboscopically obtained orbits in the Poincaré section and the Floquet modes of the transmon.

Since the rescaling of the dynamical coordinates in \cref{eq: classical Hamiltonian} does not preserve volumes in phase space, the quantization must proceed using an effective reduced Planck constant $\hbar_{\mathrm{eff}} = z = \sqrt{8E_C/E_J}$~\cite{Cohen2023Reminiscence}. The uncertainty principle then dictates that $2\pi \hbar_{\rm eff}$ is the area of phase space occupied by a single quantum state. As such, it imposes a limit to the resolution with which quantum states can probe the continuous structure of the classical phase space. While the classical limit is recovered by taking $\hbar_{\rm eff}\to 0$, we find below that classical dynamics remains a good predictor of ionization for large but finite values of $E_J/E_C$. 

The quantized Bohr-Sommerfeld orbits for oscillatory motion are the classical orbits which enclose an area equal to~\cite{Messiah2020Quantum}
\begin{equation}
\label{eq:BS_quantization}
\mathcal{A}_{i_t} = 2\pi\hbar_{\mathrm{eff}}\left(i_t+\frac{1}{2}\right),
\end{equation}
where $i_t \in \{0,1,2,3, \dots\}$ is the Bohr-Sommerfeld quantum number for the transmon. Thus, the number of Bohr-Sommerfeld states contained in a given area $\mathcal{A}$ of phase space is given by $\lfloor \mathcal{A}/2\pi\hbar_{\rm{eff}} \rfloor$. As an illustration, the orbits associated with the qubit ground and excited states are the blue and red orbits in \cref{fig:poincare_maps_2x4}. They are the orbits of the main regular region which enclose areas $\mathcal{A}_0 = \pi \hbar_{\mathrm{eff}}$ and $\mathcal{A}_1 = 3\pi\hbar_{\mathrm{eff}}$, respectively. To account for quantum fluctuations, in our analysis we extend the Bohr-Sommerfeld orbits to include the area between $\mathcal{A}_{i_t} - \pi\hbar_{\rm{eff}}$ and $\mathcal{A}_{i_t} + \pi\hbar_{\rm{eff}}$. Note that, because we are quantizing orbits obtained stroboscopically, the resulting Bohr-Sommerfeld orbits are the classical analog of the Floquet modes. We provide numerical evidence for this identification in \cref{App. Sec.: Bohr-Sommerfeld vs Floquet}.

We expect the transmon drive to induce unwanted transitions for transmon state $i_t$ whenever the Bohr-Sommerfeld orbit for that state cannot exist. Both chaotic dynamics and the presence of a $\nm{n}{m}$ resonance can prevent the existence of a given Bohr-Sommerfeld orbit. Indeed, the widening of the chaotic layer with increasing $\Tilde{\varepsilon}_t$ reduces the number of Bohr-Sommerfeld orbits supported within the main regular region. In addition, a Bohr-Sommerfeld orbit can collide with a resonance for some values of $\Tilde{\varepsilon}_t$. Moreover, as discussed further below, even a resonance appearing in the chaotic layer away from the qubit ground or excited orbits can facilitate ionization by reducing the regular area available to the qubit states.

Because it is difficult to visualize from Poincaré sections how features in phase space change with the drive amplitude, we instead plot in \cref{fig:BohrSommerfeld_drive_heatmaps}(a,b) a single cut $(\Tilde{\varphi}_t = 0, \Tilde{n}_t)$ of phase space as a function of $\Tilde{\varepsilon}_t$. All three types of motion are easily identified in this figure. Focusing on \cref{fig:BohrSommerfeld_drive_heatmaps}(a) for the moment, regular orbits correspond to the solid colored regions. The centers of each Bohr-Sommerfeld orbit are represented as pale colored lines, and the horizontal extent of the colored regions about these lines accounts for quantum fluctuations. In particular, the blue and red regions correspond to the qubit ground and excited states, respectively. The chaotic layer appears as ``noise'', and the nonlinear resonances appear as white gaps in the regular regions, with wider gaps corresponding to stronger resonances.

At large enough drive amplitudes, all Bohr-Sommerfeld orbits are eventually absorbed by the chaotic layer, and the colored regions disappear in the noise. This occurs earlier for orbits of large radius and last for the ground-state orbit. Moreover, merging of the regular region with the chaotic layer is sometimes precipitated by resonances; see, e.g., the large almost horizontal features close to $\tilde\varepsilon_t = 0.6$ and 0.9 in \cref{fig:BohrSommerfeld_drive_heatmaps}(a). Consequently, the regions where regular orbits exist can have a jagged shape reminiscent of a Christmas tree. 

Another important observation is that the distance of resonances from the center of the main regular region decreases with the drive amplitude. As a result, resonances within the main regular region collide with the Bohr-Sommerfeld orbits at some values of $\tilde{\varepsilon}_t$. For a given transmon state $i_t$, the values of $\bar{n}_r = (\omega_p/2g)^2 \Tilde{\varepsilon}_t^2$ at which the center of the Bohr-Sommerfeld orbit $i_t$ collides with a resonance or with the chaotic layer yield a set of critical photon numbers $\{n_{\mathrm{crit},i_t}^{c}\}$. These collisions are indicated by stars for $0_t$ and $1_t$. Below, we assign an uncertainty on the value of these critical photon numbers by accounting for quantum fluctuations, i.e., the horizontal extent of the colored regions. \Cref{App. Sec.: Area algorithm chaos} describes the systematic procedure that was used to build the Christmas trees, as well as to extract from them the values of $\{ n_{\mathrm{crit},i_t}^{c} \}$ and of the lower bounds that can be expected due to quantum fluctuations. 

We now discuss how the above framework enables us to understand and predict ionization thresholds for the cases of negative and positive detuning. In doing so, we connect this classical understanding to the quantum phenomenology described in the previous sections.

\subsection{Negative detuning \texorpdfstring{$\omega_q < \omega_r$}
 {\unichar{"03C9}\unichar{"1D45E} < \unichar{"03C9}\unichar{"1D45F}}}

We first consider the case of negative qubit-drive detuning; see the Poincaré sections in \cref{fig:poincare_maps_2x4}(a-d). At any finite drive amplitude $\tilde\varepsilon_t$, a chaotic layer emerges in the vicinity of the separatrix of $\tilde H_t$; see \cref{fig:poincare_maps_2x4}(b) which is obtained at the small drive amplitude $\tilde\varepsilon_t = 0.167$ corresponding to $\bar n_r = 20.6$ photons. As a result, the area of the main regular region shrinks with increasing drive amplitude at the expense of the chaotic layer. Moreover, several nonlinear resonances can be identified in the chaotic or regular regions, and we highlight in color orbits associated with a $\nm{5}{3}$ resonance, a pair of $\nm{3}{2}$ resonances, and a $\nm{7}{5}$ resonance. At this drive amplitude, neither the chaotic layer nor the nonlinear resonances disrupt the orbits associated with the qubit logical states, and we do not expect ionization to occur.

At larger drive amplitudes, the orbits of the nonlinear resonances in the main regular region shrink toward its center; see \cref{fig:BohrSommerfeld_drive_heatmaps}(a). At the drive amplitude corresponding to $\bar n_r = 185.6$, the $\nm{7}{5}$ resonance collides with the Bohr-Sommerfeld orbit associated with the qubit excited state; see \cref{fig:poincare_maps_2x4}(c), where the collision is illustrated by coloring the $\nm{7}{5}$ resonance in the color of the excited state. At that point, the qubit excited state can be unstable. As discussed further below, however, this does not necessarily result in ionization. Indeed, weak resonances occupying a small area in phase space can cross the qubit states without damage. By further increasing the drive amplitude, the orbit of the resonance shrinks and no longer coincides with that of the qubit state. Given the abundance of nonlinear resonances, the pattern of resonances crossing and then moving away from the orbits of the qubit ground and excited states is a common occurrence. This is the reason for the jagged shape of the Christmas tree in \cref{fig:BohrSommerfeld_drive_heatmaps}(a), where large resonances lead to large windows of drive amplitude for which these orbits cannot exist. This observation is in agreement with the experimental results in Ref.~\cite{Sank2016Measurement}, which showed that ``quiet spots'' with minimal measurement-induced transitions could be found at an average resonator photon number larger than the value where transitions are observed.

At still larger drive amplitudes, initializing the classical system at the energy corresponding to the qubit excited state does not lead to a regular orbit but rather fills the chaotic layer; see \cref{fig:poincare_maps_2x4}(d) obtained for $\bar n_r = 287.6$. Thus, at that drive amplitude, the excited state is engulfed by the chaotic layer and is unstable. At yet larger drive amplitudes (not shown), the ground state eventually suffers the same fate and becomes unstable. It is interesting to note the presence in \cref{fig:poincare_maps_2x4}(d) of a pair of $\nm{3}{2}$ resonances near the inner edge of the chaotic layer. Because to each resonance is associated a new separatrix around which chaos emerges, the presence of these $\nm{3}{2}$ resonances further reduces the phase space available to the orbit associated with the qubit excited state, thereby hastening its merger with the chaotic layer. We note that the collisions of $\nm{n}{m}$ resonances with the Bohr-Sommerfeld orbits associated with the qubit states are the classical analogs of the multiphoton resonances discussed in \cref{subsec: fully quantum negative detuning} and \cref{subsec: floquet negative detuning}, where ionization of the qubit states is delayed until the drive pushes higher-energy dressed states into resonance with the computational subspace.

\begin{figure}[t!]
 \centering
 \includegraphics[width=\linewidth]{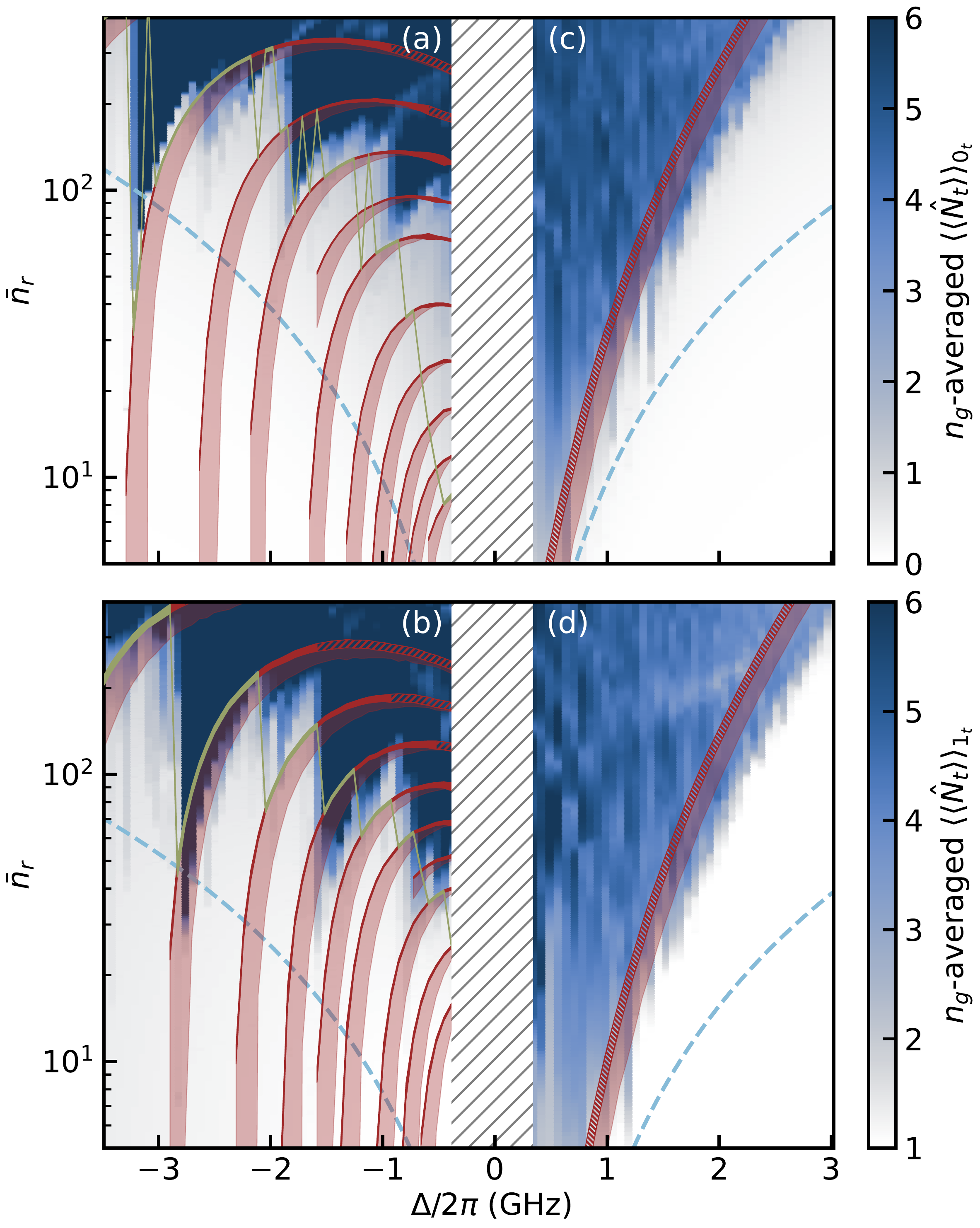}
 \caption{Bohr-Sommerfeld critical photon numbers $\{n_{\mathrm{crit},i_t}^{c}\}$ (red lines) and gate charge average of the period-averaged transmon population $\langle\langle \hat{N}_t \rangle\rangle_{i_t}$ defined in \cref{eq: period-averaged transmon excitation in each Floquet mode} (colored background) as a function of detuning $\Delta$ and photon number $\bar n_r$. The top corresponds to the qubit ground state for (a) negative detuning and (c) positive detuning, and the bottom corresponds to the qubit excited state for (b) negative detuning and (d) positive detuning. The detuning is varied by changing the drive frequency. Note the change in color scale between the top and bottom. The hatched red lines indicate the critical photon numbers due to the chaotic layer, while the solid red lines indicate the critical photon numbers due to a resonance. The width of the solid red lines is proportional to the phase-space area of the resonance, while that of the hatched red lines is chosen to correspond to an area of $2\pi\hbar_{\rm eff}$. The shaded red areas give the lower-bound uncertainty due to quantum fluctuations around the Bohr-Sommerfeld orbit. The dashed blue lines are the Jaynes-Cummings-like critical photon numbers $n_{{\rm crit}, i_t}$ defined in \cref{App. Sec.:dispersive_ncrit} for each state. The hatched white regions indicate regions where the dispersive approximation does not hold and for which critical photon numbers are not calculated. In (a) and (b), the olive lines indicate the smallest critical photon number for which the phase-space area of the corresponding resonance is larger than $0.05 \times 2\pi \hbar_{\rm eff}$ and $0.10 \times 2\pi \hbar_{\rm eff}$, respectively. For both classical and Floquet theory, the vertical axis is produced by varying $\varepsilon_t/2\pi$ in increments of $\SI{10}{MHz}$.}
 \label{fig:Floquet_Nt_CAncrits} 
\end{figure}

Building on these observations, \cref{fig:Floquet_Nt_CAncrits}(a,b) show the critical photon numbers $\{n_{\mathrm{crit},i_t}^{c}\}$ as a function of detuning $\Delta$. The solid red lines indicate a collision of the orbit associated with the qubit (a) ground or (b) excited state with a resonance, while the hatched red lines rather indicate a collision with the chaotic layer. At $\Delta/2\pi = -\SI{1.202}{GHz}$, these lines correspond to the stars in \cref{fig:BohrSommerfeld_drive_heatmaps}(a). While chaos is expected to always induce ionization, this is not the case for all collisions with resonances. This is because many resonances occur at low photon numbers and are, thus, too weak to induce ionization. To account for this, the width of the solid red lines is proportional to the phase-space area of the resonance colliding with the qubit state orbit, which is, in turn, proportional to the resonance's strength. Using this information, the olive lines indicate the smallest critical photon number for which the phase-space area of the corresponding resonance is larger than $0.05 \times 2\pi \hbar_{\rm eff}$ in \cref{fig:Floquet_Nt_CAncrits}(a) and $0.10 \times 2\pi \hbar_{\rm eff}$ in \cref{fig:Floquet_Nt_CAncrits}(b). These values are free parameters and were chosen to best fit the classical model to the Floquet analysis; see the discussion below. The occasionally jagged behavior of that line in \cref{fig:Floquet_Nt_CAncrits}(a) occurs when the area of the relevant resonances becomes comparable to the resolution with which we compute phase-space areas.

These Bohr-Sommerfeld critical photon numbers are compared to the period-averaged transmon population $\langle\langle \hat{N}_t \rangle\rangle_{i_t}$ (colored background), a quantity which was shown to be a good indicator of transmon ionization in \cref{sec: Floquet with just transmon}. We average the results over $n_g$ to remove the dependence on this parameter. In both panels, we observe abrupt changes in $\langle\langle \hat{N}_t \rangle\rangle_{i_t}$ near the photon numbers where classical resonances occur. The agreement is excellent when discarding weak resonances which are not expected to have an effect on the qubit; see the olive lines. That some resonances are not relevant to transmon ionization is consistent with the discussion in \cref{sec: Floquet with just transmon}, where it was observed that weak resonances between Floquet modes do not cause ionization because they can be traversed diabatically. Other correspondences between the chaotic behavior of the driven classical pendulum and Floquet quasienergies of the transmon are discussed in Ref.~\cite{Cohen2023Reminiscence}.

\subsection{Positive detuning \texorpdfstring{$\omega_q > \omega_r$}
 {\unichar{"03C9}\unichar{"1D45E} > \unichar{"03C9}\unichar{"1D45F}}}
\label{subsec:classical_positive_det}

We now turn to the case of positive detuning; see \cref{fig:poincare_maps_2x4}(e-h). As at negative detuning, a chaotic layer develops when the drive amplitude is made finite; see \cref{fig:poincare_maps_2x4}(f) obtained for $\bar n_r = 0.63$ photons. The most dramatic feature is, however, the appearance of a large $\nm{1}{1}$ resonance inside the separatrix of $\tilde H_t$. In \cref{fig:BohrSommerfeld_drive_heatmaps}(b), the dome-shaped feature centered at $\tilde n_t = 0$ at low drive amplitude corresponds to the main regular region, while the Christmas-tree-like feature centered at $\tilde n_t \sim 1.6$ corresponds to the resonance. This resonance results from the negative anharmonicity of the transmon, which at positive detuning allows for $\tilde \omega(\tilde I) = \tilde \omega_d$ to be satisfied for some $\tilde I$. This is the classical equivalent of the breakdown of the dispersive approximation at zero photon number due to a one-photon qubit-resonator resonance discussed in \cref{subsec: fully quantum positive detuning}. 

While this resonance typically does not collide with the Bohr-Sommerfeld orbits associated with the qubit ground and excited states, its presence rapidly restricts the phase space available to the main regular region. This is exacerbated by the presence of a new separatrix surrounding the $\nm{1}{1}$ resonance and about which a chaotic layer develops; see \cref{fig:poincare_maps_2x4}(g) obtained at $\bar n_r = 4.88$ photons and the V-shaped region between the main regular region and the resonance in \cref{fig:BohrSommerfeld_drive_heatmaps}(b). Increasing further the drive amplitude, the qubit excited state is engulfed by the chaotic layer and becomes unstable; see \cref{fig:poincare_maps_2x4}(h) obtained at $\bar n_r = 17.02$ photons. This is followed by the ground state also entering the chaotic layer at a still larger photon number.

The area of the $\nm{1}{1}$ resonance is large enough ($> 2\pi \hbar_{\rm eff}$) to support its own Bohr-Sommerfeld orbits; see \cref{App. Sec.: Bohr-Sommerfeld vs Floquet}. In \cref{fig:BohrSommerfeld_drive_heatmaps}(b,d), these orbits share the same color code as the distinct orbits of the main regular region because they enclose the same areas. These orbits are located in phase space within the separatrix of $\tilde H_t$ and, thus, have an energy that is below the top of the cosine potential well of the transmon. This is in direct analogy with the branch bunching about a state $i_t^\star$ located below the top of the well discussed in \cref{subsec: fully quantum positive detuning,subsec: floquet positive detuning}. Moreover, additional $\nm{n}{m}$ resonances appear within the main regular region; see \cref{fig:BohrSommerfeld_drive_heatmaps}(b). However, these are small and are usually not relevant compared to the large $\nm{1}{1}$ resonance discussed above. Only at relatively large positive detuning does the critical photon number associated with the $\nm{1}{1}$ resonance become large enough to allow other $\nm{n}{m}$ resonances to become relevant for ionization (not shown).

\cref{fig:Floquet_Nt_CAncrits}(c,d) show the critical photon numbers $\{n_{\mathrm{crit},i_t}^{c}\}$ 
as a function of positive detuning $\Delta$ and average photon number $\bar n_r$, with \cref{fig:Floquet_Nt_CAncrits}(c) corresponding to the ground state and \cref{fig:Floquet_Nt_CAncrits}(d) corresponding to the excited state. Because the small $\nm{n}{m}$ resonances crossing the qubit-state orbits are of no consequence for this range of detuning, the critical photon number here corresponds to the point where the qubit states are engulfed by the chaotic layer (hatched red lines). For most detunings, the presence of the large $\nm{1}{1}$ resonance results in a much smaller critical photon number than at negative detuning. Also in contrast with the case of negative detuning, the Bohr-Sommerfeld critical photon number monotonically increases with the magnitude of the detuning, since the $\nm{1}{1}$ resonance dominates the ionization mechanism.

The critical photon number is compared with the gate charge average of $\langle\langle \hat{N}_t \rangle\rangle_{i_t}$ (colored background). Although the classical predictions tend to slightly overestimate the critical photon number obtained from Floquet theory, we again observe a good agreement between the two approaches over a large range of detunings and photon numbers. Some features that are not captured are the oscillations in the critical photon number at small detuning; see \cref{fig:Google_results} for a clearer view of the oscillating features. As discussed in \cref{subsec: fully quantum positive detuning} and \cref{subsec: floquet positive detuning}, these arise from quantum tunneling between pairs of states and, thus, cannot be captured within a classical model. In broad strokes, however, the classical model of the driven nonlinear pendulum captures the essence of transmon ionization remarkably well.
 
\section{Comparison to experimental results \label{sec:prediction_experiment}}

\begin{figure}[t]
 \centering
 \includegraphics[width=\linewidth]{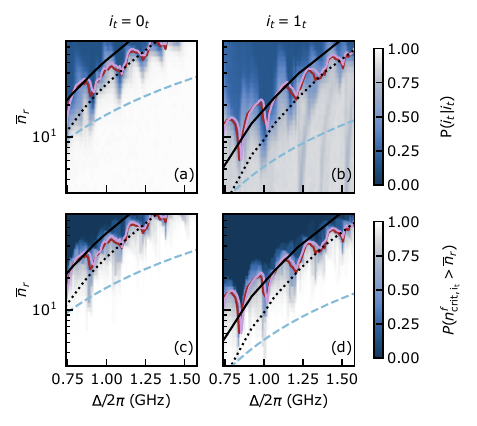}
 \caption{
 Critical photon numbers of our three models compared to the experimental results in Ref.~\cite{Khezri2023Measurement}. The red line shows our prediction of $\bar{n}_{{\rm crit}, i_t}^q$ using the branch analysis of the full transmon-resonator system. The pink line shows $\bar{n}_{{\rm crit}, i_t}^f$ obtained from the Floquet branch analysis. The solid black line shows the prediction from the classical dynamics, with a lower bound given by the dotted black line. The dashed blue line shows Jaynes-Cummings-like critical photon numbers defined in \cref{App. Sec.:dispersive_ncrit} for each state. In (a) and (b), the heat maps show experimental results for the ground state ($i_t=0_t$) and for the excited state ($i_t=1_t$), respectively. For a fixed detuning between the tunable transmon frequency and the resonator frequency, the experiment consists of preparing the transmon qubit in $\ket{0_t}$ or $\ket{1_t}$, driving the resonator at the conditional dressed frequencies $\omega_{r,0_t}$ or $\omega_{r,1_t}$ to populate it with $\bar n_r$ photons on average, waiting for the resonator to decay back to the vacuum, and finally reading out the qubit to extract the conditional probabilities $P(i_t|i_t)$ of finding the outcome $i_t$ given that $\ket{i_t}$ was initially prepared. The long vertical features in (b) at very low photon numbers and for $\Delta/2\pi\gtrsim\SI{1.25}{GHz}$ are attributed to the presence of two-level defects~\cite{Khezri2023Measurement}. In (c) and (d), the heat maps show the cumulative distribution function $P(n_{{\rm crit}, i_t}^f>\overline{n}_r)$ over realizations of gate charge for the ground state ($i_t=0_t$) and for the excited state ($i_t=1_t$), respectively. The lines plotted in (c) and (d) are the same as the lines plotted in (a) and (b), respectively. }
 \label{fig:Google_results}
\end{figure}

We now compare the critical photon numbers extracted from the quantum transmon-resonator, Floquet, and classical models to experimental data from Ref.~\cite{Khezri2023Measurement}. This experiment consists in preparing the qubit in its ground or first excited state, filling the resonator with $\bar n_r$ photons for a controlled amount of time, and then performing a QND readout of the qubit state. The colored background in \cref{fig:Google_results}(a,b) shows the conditional probability $P(i_t|i_t)$ for the qubit to be measured in the state it was prepared in as a function of detuning and average resonator photon number. \cref{fig:Google_results}(a) is for the ground state and \cref{fig:Google_results}(b) is for the excited state. A white background indicates that the interaction with the $\bar n_r$ photons did not cause qubit leakage, while a dark background indicates leakage. In both cases, sharp dips are observed as a function of detuning. In that experiment, the readout resonator has a bare frequency $\omega_r/2\pi=\SI{4.75}{GHz}$, the coupling constant is $g/2\pi=\SI{91}{MHz}$, the charging energy is $E_C/2\pi = \SI{194}{MHz}$, and the qubit frequency $\omega_q/2\pi$ can be tuned from $5.5$ to $\SI{6.34}{GHz}$. In all cases, this corresponds to the regime of positive detuning. 

In Ref.~\cite{Khezri2023Measurement}, a semiclassical model predicting the positions and amplitudes of these dips is presented. Here, we compare the experimental results to the critical photon number obtained from our gate-charge-averaged fully quantum transmon-resonator (pink line) and semiclassical driven transmon (red line) models. The agreement with experimental results is excellent for both models and qubit initial states. In particular, the theoretical values capture the full detuning dependence of the critical photon number, including the presence of dips in $\bar n_r$. These dips correspond to the pairwise multiphoton resonances which are expected to precipitate ionization at positive detuning; see \cref{subsec: fully quantum positive detuning}. Notably, the theory captures not only the position of these dips, but also their height and width. Moreover, as expected from the discussion in the previous sections, ionization typically occurs earlier for the qubit excited state than for the ground state. Considering the experiment's slow timescale relative to gate charge fluctuations, we also show the full cumulative distribution function $P(\bar{n}_{{\rm crit}, i_t}^f > \bar{n}_r)$ of the Floquet critical photon numbers $\bar{n}_{{\rm crit}, i_t}^f$ over many iterations of gate charge as the colored background in \cref{fig:Google_results}(c,d). The gate charge variance of this distribution is in qualitative agreement with the critical photon number variance observed in the experiment, emphasizing the necessity of including gate charge in these predictions. Because it predicts the full dependence of the ionization threshold with detuning, including regions where $\bar{n}_{{\rm crit}, 0_t}$ and $\bar{n}_{{\rm crit}, 1_t}$ are large in between resonances, the framework presented here could be exploited to maximize readout performance.

The prediction from the classical model (solid black lines) follows the general trend of the critical photon number as a function of detuning. The dotted black lines are a lower bound to these predictions that accounts for quantum fluctuations. Because the classical model does not capture the pairwise multiphoton resonances responsible for accelerating ionization, the classical predictions do not capture the dips in critical photon number versus detuning. Yet, it is remarkable that, as long as the full cosine potential is accounted for, excellent estimates of the photon number threshold for ionization that match experimental results can be obtained simply by solving Hamilton's equations of motion.

\section{Conclusion and outlook} \label{sec:conclusions}

We have introduced a comprehensive framework describing the physical origin of transmon ionization. Our study is based on three complementary levels of description of the qubit-resonator system used for dispersive qubit readout in circuit QED. A first approach treats the qubit and the resonator as coupled quantum systems. A second approach simplifies the situation by replacing the driven resonator by a classical drive acting directly on the transmon. A third approach considers an entirely classical problem. All three approaches have in common that they rely on the full cosine potential of the transmon's Josephson junction. The first two descriptions identify multiphoton resonances as the mechanism responsible for ionization. The positions of these resonances can be easily located and estimates of critical photon numbers can be efficiently obtained numerically with both methods, leading to predictions which are in agreement with each other and with recent experimental results. These approaches also highlight the strong dependence of the critical photon numbers on gate charge, even deep in the transmon regime. The third level of description provides a simple interpretation of ionization based on the motion of a classical driven pendulum and its chaotic behavior. Although this method misses some features captured by the quantum models, the critical photon numbers it predicts are in qualitative agreement with the quantum predictions and with experimental results. It is remarkable that classical dynamics of the driven transmon is sufficient to capture the physics of ionization.

Although we used fixed sets of parameters throughout this work, we verified that the conclusions we reached remain general. While, e.g., the specific multiphoton resonances precipitating ionization may shift with changes in parameters, the dependence of the critical photon numbers $\bar{n}_{{\rm crit}, i_t}^q$ and $\bar{n}_{{\rm crit}, i_t}^f$ on detuning preserves the same features.

Having identified the root causes of ionization, it remains to be seen how to exploit our findings to increase readout speed and fidelity. Our work, however, already offers several general strategies to keep in mind when designing qubits. For example, one should exploit regions of detunings located between the dominant multiphoton resonances, where the critical photon number can be large. In addition, the increase in critical photon number with increasing absolute detuning $|\Delta|$ (see, e.g., \cref{fig:comparison_quantum_floquet}) indicates the potential of working at large detuning to improve qubit readout. While the trade-off between the reduced dispersive shift and the increased critical photon number in this regime has not yet been extensively investigated, recent experimental findings lend support to this hypothesis~\cite{DevoretPersonal2024}. Moreover, because the speed at which resonances are traversed affects the probability of ionization occurring, optimizing the readout pulse shape with an understanding of the position of resonances could prove to be beneficial.
We hope that the analyses presented in this work will facilitate the experimental exploration of parameter regions wherein transmon ionization occurs at large photon numbers.

Beyond dispersive readout, many experiments in circuit QED contain the basic ingredients that are at the core of ionization, namely, strong drives and weak anharmonicities. These include, but are not limited to, other types of qubits \cite{Manucharyan2009-Fluxonium, Hyyppa2022-Unimon, Gusenkova2021, Bista2024Effects}, readout schemes \cite{Didier2015_Longitudinal,Chapple_2024_Longitudinal}, parametric gates \cite{Bertet2006,Niskanen2006,Liu2007,Beaudoin2012,McKay2016}, and quantum state stabilization \cite{Puri2017_Kerr_Cat, Frattini2024_Sqeezed_Kerr, Chavez-Carlos2023_Melting_Chaos}. The tools introduced in this work can easily be applied to these scenarios. These tools are simple to interpret, have an easy construction, are easy to implement, and are computationally efficient. Thus, we anticipate the framework introduced in this work to become an integral part of the theoretical analysis and support of current and future circuit QED experiments. 

%%%%%%%%%%%%%%%%%%%%%%%%%%%%%%%%
%%%%%%% Acknowledgements %%%%%%%
%%%%%%%%%%%%%%%%%%%%%%%%%%%%%%%%

\section*{Acknowledgements}
The authors are grateful to Ross Shillito and Pierre-Antoine Graham for helpful discussions. This work is supported by a collaboration between the U.S. Department of Energy and other agencies. This material is based upon work supported by the U.S. Department of Energy, Office of Science, National Quantum Information Science Research Centers, Quantum Systems Accelerator. Additional support is acknowledged from the Natural Sciences and Engineering Research Council, the Canada First Research Excellence Fund, the Minist\`ere de l’\'Economie et de l’Innovation du Qu\'ebec, and Fonds de recherche du Qu\'ebec - Nature et technologies.

%%%%%%%%%%%%%%%%%%%%%%%%%%%%%%%%
%%%%%%%%% Appendix %%%%%%%%%%%
%%%%%%%%%%%%%%%%%%%%%%%%%%%%%%%%

\appendix

\section{Jaynes-Cummings-like critical photon numbers}
\label{App. Sec.:dispersive_ncrit}

In the main text, we compare the various critical photon numbers defined in our work with the critical photon numbers $n_{{\rm crit}, i_t}$ that define the dispersive limit. Here, we summarize how these critical photon numbers are calculated.

The commonly used Jaynes-Cummings critical photon number for the qubit transition is defined as $n_{\rm crit} = |(\omega_{0_t 1_t}-\omega_r)/2g_{0_t 1_t}|^2$~\cite{Blais2004Cavity, Blais2021Circuit}. This is the photon number for which the interaction term in \cref{eq:transmon_resonator_hamiltonian} becomes resonant with the qubit transition. However, this definition is insufficient to capture the breakdown of the dispersive regime in a multilevel system like the transmon. It is instead necessary to define a critical photon number for every transition~\cite{Danjou2019Optimal, Blais2021Circuit}. To do so, we simply transpose the definition for the qubit transition to all other transitions:
\begin{align}
\begin{aligned}
  n_{{\rm crit}, k_t l_t} \equiv \left|\frac{\omega_{k_t l_t} - \omega_r}{2g_{k_t l_t}}\right|^2. \label{eq:dispersive_ncrit}
\end{aligned}
\end{align}
Note that this definition naturally includes counterrotating transitions as the case where $\omega_{k_t l_t} \sim -\omega_r$. We define the Jaynes-Cummings-like critical photon number for an initial transmon state $|i_t\rangle$ as the smallest of the $n_{{\rm crit}, k_t l_t}$ that involves the initial state:
\begin{align}
 n_{{\rm crit}, i_t} = \min_{k_t l_t} \{n_{{\rm crit}, k_t l_t} | k_t = i_t \text{ or } l_t = i_t\}.
\end{align}
Because of the dipolelike selection rules of the transmon regime, the limiting transitions for the qubit states are always $0_t \leftrightarrow 1_t$ and $1_t \leftrightarrow 2_t$. As a result, $n_{{\rm crit}, 0_t}$ and $n_{{\rm crit}, 1_t}$ are essentially independent of the gate charge $n_g$. \Cref{eq:dispersive_ncrit} gives similar predictions as other definitions of the state-dependent critical photon numbers for the transmon~\cite{Blais2021Circuit} but has the advantage of being applicable to any multilevel system.

\section{Nonperturbative corrections to the branches at negative detuning} \label{App. Sec.: Non-perturbative corrections}

In \cref{subsec: fully quantum negative detuning}, we argue that nonperturbative corrections to the approximate eigenstates $|\dashover{i_t,n_r}\rangle \equiv e^{-\hat{S}}|i_t, n_r\rangle \approx |\overline{i_t,n_r}\rangle$ obtained via Schrieffer-Wolff perturbation theory are ultimately the cause of branch swapping. Here, we discuss at length the mechanism through which these nonperturbative corrections occur.

As an illustrative example, let us consider the swapping of branches $B_{1_t}$ and $B_{7_t}$ which occurs at roughly 84 photons in \cref{fig: fully quantum negative detuning}. If we were to consider only the bare transmon energies, swapping between these branches would be unexpected. Indeed, even if one considers the possibility of a multiphoton process connecting such states, the detuning between them is still large. For instance, $(E_{7_t} - E_{1_t} - 4 \omega_r)/2\pi \approx \SI{1.98}{GHz} \gg g / 2\pi$ is the ``most-resonant" multiphoton transition for states near the top of the cosine potential. Moreover, the interaction cannot directly add or remove four photons from the resonator, and any such process that appears within perturbation theory should be small. Thus, it is at this stage not evident why the swapping should occur between these branches.

To address how the requisite term mediating this process arises, we must keep more terms in the infinite-order expansion than in \cref{eq:SW_H_0}. To that end, we recall that the standard approach~\cite{Koch2007ChargeInsensitive, Schrieffer1966Relation} dictates that by choosing
\begin{align}
 \hat{S} = 
 -i (
 \hat{S}_t \ahat^\dagger
 +
 \ahat \hat{S}_t^\dagger
 )
\end{align}
with 
\begin{align}
 \hat{S}_t
 =
 \sum_{i_t, j_t = 0}^{\infty}
 \frac{g_{i_t j_t} }{\omega_{i_t j_t}-\omega_r}
 |i_t\rangle \langle j_t |,
\end{align}
we obtain
\begin{align} \nonumber 
 &e^{\hat{S}}\hat{H}_{tr} e^{-\hat{S}}
 =
 \omega_r \ahat^\dagger \ahat + \hat{H}_t
 +
 \ahat^\dagger \ahat 
 \sum_{i_t, j_t = 0}^{\infty}
 \chi_{i_t j_t}
 |i_t\rangle \langle j_t |
 \\ \nonumber
 &+
 \left( 
 \ahat^\dagger \ahat^\dagger 
 \sum_{i_t, j_t= 0}^\infty
 \eta_{i_t j_t}
 |i_t\rangle \langle j_t|
 +
 \mathrm{H.c.}
 \right)
 +
 \sum_{i_t, j_t = 0}^\infty
 \Lambda_{i_t j_t} |i_t \rangle \langle j_t |
 \\ \label{app.eq:H_SW}
 &
 +
 O(g^3),
\end{align}
where the matrix elements are
\begin{align}
 \chi_{i_t j_t}
 &=
 \sum_{k_t = 0}^\infty
 g_{i_t k_t}g_{k_t j_t}
 \left[ 
 \frac{ \omega_{i_t k_t}}
 {\omega_r^2 - (\omega_{i_t k_t})^2}
 +
 \frac{ \omega_{j_t k_t}}
 {\omega_r^2 - (\omega_{j_t k_t})^2}
 \right],
 \\
 \eta_{i_t j_t}
 &=
 \frac{1}{2}
 \sum_{k_t = 0 }^\infty
 g_{i_t k_t} g_{k_t j_t}
 \left( 
 \frac{1}
 {\omega_r - \omega_{k_t j_t} }
 -
 \frac{1}
 {\omega_r + \omega_{k_t i_t}}
 \right),
 \\
 \Lambda_{i_t j_t}
 &=
 -\frac{1}{2}
 \sum_{k_t = 0}^\infty
 g_{i_t k_t}g_{k_t j_t}
 \left(
 \frac{1}{\omega_r - \omega_{k_t j_t}}
 +
 \frac{1}{\omega_r - \omega_{k_t i_t}}
 \right),
\end{align}
and we drop all terms of the order of $g^3$ and higher. The dispersive $\chi$ shifts and Lamb shifts discussed in the main text are the diagonal parts $\chi_{i_t} \equiv \chi_{i_t i_t}$ and $\Lambda_{i_t} \equiv \Lambda_{i_t i_t}$. Note that we use a different notation than that presented in Ref.~\cite{Koch2007ChargeInsensitive}.
We also note that \textcite{Xiao2023Diagrammatic} introduce a diagrammatic approach to obtain a high-order effective Hamiltonian for a driven transmon in the absence of a resonator.

By keeping the diagonal terms in \cref{app.eq:H_SW}, we recover \cref{eq:SW_H_0} and, thus, correctly obtain the dispersion to the order $g^2$. However, a salient feature of \cref{app.eq:H_SW} is that, to the same order in the coupling, there are terms like $\ahat^\dagger \ahat ^\dagger |j_t\rangle \langle k_t|$ that create pairs of photons in the resonator~\cite{Danjou2019Optimal}. These terms can also cause transitions between any pair of approximate eigenstates with an amplitude proportional to $\eta_{i_t j_t}$. If we were to explicitly keep track of terms to higher order in $g$, we would similarly conclude that to fourth order there appears an analogous term $\ahat^\dagger \ahat^\dagger \ahat^\dagger \ahat^\dagger |j_t\rangle \langle k_t|$. Nevertheless, it would seem that this term is irrelevant, considering that it is of fourth order in a nominally small parameter. 

To ensure that such terms can safely be dropped, however, we would have to verify that, in this new basis of approximate eigenstates, the usual criteria for the validity of perturbation theory is satisfied. As we emphasize in the main text, these states are ac-Stark shifted. This opens up the possibility that these new approximate eigenstates can have a very small detuning, thus rendering seemingly small off-diagonal terms like $\ahat^\dagger \ahat^\dagger \ahat^\dagger \ahat^\dagger |j_t\rangle \langle k_t|$ relevant. This necessarily leads to nonperturbative corrections to the eigenstates.

For instance, focusing again for concreteness on the branches $B_{1_t}$ and $B_{7_t}$ in \cref{fig: fully quantum negative detuning}, suppose we had computed the perturbative corrections to the bare states $|1_t, n_r\rangle$ and $|7_t, m_r\rangle$ to some order in $g$ using higher-order Schrieffer-Wolff perturbation theory. These approximate eigenstates $|\dashover{1_t,m_r}\rangle$ and $|\dashover{7_t, n_r}\rangle$ have energies $\langle \dashover{1_t,m_r}| H_{tr} |\dashover{1_t,m_r} \rangle $ and $\langle \dashover{7_t,n_r}| H_{tr} |\dashover{7_t,n_r} \rangle $, respectively. They also have a residual coupling with an interaction strength $\langle \dashover{1_t,m_r}| H_{tr} |\dashover{7_t,n_r} \rangle $, which is necessarily proportional to a nonzero power of $g$. 

As stated above, the most-resonant transition between the $1_t$ and $7_t$ transmon states involves four photons. Thus, assuming that the coupling with any other approximate eigenstate is irrelevant, we can project onto the two-dimensional subspace spanned by $|\dashover{1_t,n_r+4} \rangle$ and $|\dashover{7_t,n_r} \rangle$ for each $n_r$ and conclude that the eigenvalues and eigenvectors of the two-dimensional matrix 
\begin{align}
 \begin{pmatrix}
  \langle \dashover{1_t,n_r+4}| H_{tr} |\dashover{1_t,n_r+4} \rangle & \langle \dashover{1_t,n_r+4}| H_{tr} |\dashover{7_t,n_r} \rangle 
  \\
  \langle \dashover{7_t,n_r}| H_{tr} |\dashover{1_t,n_r+4} \rangle 
  &
  \langle \dashover{7_t,n_r}| H_{tr} |\dashover{7_t,n_r} \rangle
 \end{pmatrix}
\end{align}
control the hybridization between these approximate eigenstates. Thinking of $n_r$ as an external controllable parameter, we would observe an avoided crossing when the condition 
\begin{align}\label{app.eq:SW_Resonance}
 \frac{
  |\langle \dashover{1_t,n_r+4} | 
 \hat{H}_{tr}
 |\dashover{7_t,n_r} \rangle|
 }
 {
 \left| \langle \dashover{7_t,n_r} | 
 \hat{H}_{tr}
 |\dashover{7_t,n_r} \rangle
 -
 \langle \dashover{1_t,n_r+4} | 
 \hat{H}_{tr}
 |\dashover{1_t,n_r+4} \rangle
 \right|
 }
 \gtrsim
 1
\end{align}
is met. Had we gone to high-enough order in $g$, we would have found that \cref{app.eq:SW_Resonance} is satisfied for $n_r \approx 84$.

Thus, we see that the branch analysis classifies states in a way that is analogous to adiabatically tracking the energies of each state. For instance, before the crossing the off-diagonal elements are irrelevant, and the branch analysis classifies $|\overline{1_t, n_r}\rangle \approx |\dashover{1_t,n_r} \rangle $ and $|\overline{7_t, n_r}\rangle \approx |\dashover{7_t,n_r} \rangle $. After the crossing, the characters of the states are switched, and we have $|\overline{1_t,n_r} \rangle \approx |\dashover{7_t,n_r - 4} \rangle$ and $|\overline{7_t,n_r} \rangle \approx |\dashover{1_t,n_r+4} \rangle$. Near the crossing itself, however, the eigenstates are a highly entangled superposition $ \approx |\dashover{1_t,n_r+4} \rangle \pm |\dashover{7_t,n_r}\rangle$, and there is an accompanying avoided crossing in the modular spectrum. We stress that there is no way to perturbatively connect the entangled states $|\dashover{1_t,n_r+4} \rangle \pm |\dashover{7_t,n_r}\rangle$ to the approximate eigenstates $|\dashover{1_t,n_r+4} \rangle$ and $|\dashover{7_t,n_r} \rangle$, and so by definition they could not have been obtained via perturbation theory in the residual coupling between these states. This highlights the nonperturbative nature of branch swapping.

\begin{figure}[t]
 \centering
 \includegraphics[width=\linewidth]{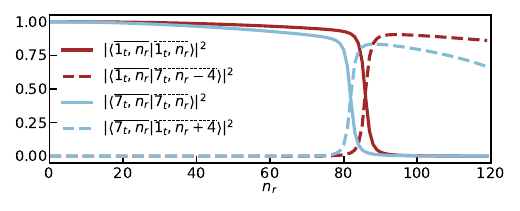}
 \caption{Overlaps of the eigenstates of branches $B_{1_t}$ and $B_{7_t}$ with the corresponding perturbative approximations $|\dashover{i_t,n_r}\rangle$ obtained via Schrieffer-Wolff perturbation theory [\cref{app.eq:SW_states}]. The parameters are the same as in \cref{fig: fully quantum negative detuning}. Before the branch swapping at approximately 84 photons, the eigenstates in each branch are well approximated by their perturbative counterpart. Near the swapping, perturbation theory fails and the eigenstates are in a roughly equal superposition of the perturbative eigenstates. For larger photon numbers, the character of the eigenstates switches. Deviation of the overlaps from unity beyond the swapping point is a consequence of the finite-order approximation for the $|\dashover{i_t,n_r}\rangle$, which fails at large photon numbers.
 }
 \label{fig:Non_perturbative_overlaps}
\end{figure}

We now wish to confirm the validity of this interpretation by comparing the overlaps of the approximate eigenstates with the relevant numerically obtained eigenstates; see \cref{fig:Non_perturbative_overlaps}. It is important to recall that the coupling scales as $g\sqrt{n_r}$ such that the validity of perturbation theory worsens with increasing photon number. As a result, the overlaps between the numerically exact eigenstates and their perturbative counterparts decrease as $n_r$ gets large. To try and remove this trivial decrease in the overlaps, which should be contrasted with the nonperturbative corrections due to an avoided crossing, we now go to one order higher in perturbation theory than in \cref{app.eq:H_SW}. By defining
\begin{align}\label{app.eq:SW_states}
 |\dashover{i_t, n_r}\rangle
 \equiv
 e^{-\hat{S}}
 e^{-\hat{T}}
 |i_t, n_r \rangle,
\end{align}
where $\hat{T}$ is proportional to $g^2$, we can eliminate the off-diagonal terms to that order in \cref{app.eq:H_SW}. Just like $\hat{S}$ is linear in $\hat{a}$ and $\hat{a}^\dag$ to eliminate the first-order term in $g$, $\hat{T}$ is quadratic in $\hat{a}$ and $\hat{a}^\dag$ to eliminate the second-order terms in $g$. Its full explicit form is 
\begin{align}
 \hat T&=\hat A_t^{(2)}+\hat B_t^{(2)}\hat a^\dag \hat a+\hat C_t^{(2)}\hat a \hat a+\hat D_t^{(2)}\hat a^\dag \hat a^\dag,
\end{align}
where
\begin{align}
 \hat A_t^{(2)}&=-\frac{g}{2}\sum_{i_t\neq j_t}\frac{\ket{i_t}\bra{i_t}(\hat n_t\hat S_t + \hat S_t^{\dag} \hat n_t)\ket{j_t}\bra{j_t}}{\omega_{i_tj_t}},\\
 \hat B_t^{(2)}&=\frac{g}{2}\sum_{i_t\neq j_t}\frac{\ket{i_t}\bra{i_t}\big[\hat n_t,\hat S_t^{\dag}-\hat S_t\big]\ket{j_t}\bra{j_t}}{\omega_{i_tj_t}},\\
 \hat C_t^{(2)}&=\frac{g}{2}\sum_{i_t, j_t}\frac{\ket{i_t}\bra{i_t}\big[\hat S_t^{\dag},\hat n_t\big]\ket{j_t}\bra{j_t}}{\omega_{i_tj_t}+2\omega_r},\\
 \hat D_t^{(2)}&=\frac{g}{2}\sum_{i_t, j_t}\frac{\ket{i_t}\bra{i_t}\big[\hat n_t,\hat S_t\big]\ket{j_t}\bra{j_t}}{\omega_{i_tj_t}-2\omega_r}.
\end{align}
At low photon numbers $n_r < 84$, we see in \cref{fig:Non_perturbative_overlaps} that $|\langle \overline{1_t, n_r}|\dashover{1_t, n_r}\rangle|^2 \approx |\langle \overline{7_t, n_r}|\dashover{7_t, n_r}\rangle|^2 \approx 1$, indicating the validity of the dispersive approximation. For $n_r > 84$, the overlaps $|\langle \overline{1_t, n_r}|\dashover{7_t, n_r-4}\rangle|^2$ and $|\langle \overline{7_t, n_r}|\dashover{1_t, n_r+4}\rangle|^2$ rise and approach unity, indicating branch swapping and the validity of perturbation theory. Near $n_r \approx 84$, however, all of these overlaps are roughly a half; if we were to perform higher-order Schrieffer-Wolff perturbation theory, we could not recover a nearly flat line of order unity at this value of $n_r$, since the validity of this perturbative expansion is contingent on the left-hand side of \cref{app.eq:SW_Resonance} being small. 

The fact that the overlaps significantly deviate from unity on each side of the branch swapping is a consequence of using a finite-order expansion for the approximate eigenstates. Indeed, such an expansion becomes less accurate at large $n_r$, since the transmon-resonator interaction strength scales with photon number. The overlaps could be made arbitrarily close to unity before and after the crossing by going to even higher order in perturbation theory. At the crossing itself, however, the observed branch swapping is always nonperturbative in the coupling $g$. 

We emphasize that we focused on $B_{1_t}$ and $B_{7_t}$ since the branch analysis had already identified that these two branches had swapped. Without the branch analysis, we would need to verify the resonance condition [\cref{app.eq:SW_Resonance}], for every pair of branches. Furthermore, we would need to ensure that the perturbative energies $\langle \dashover{i_t,n_r} | \hat{H}_{tr} |\dashover{i_t,n_r} \rangle$ are accurate. At higher photon numbers, this means going to higher order in perturbation theory, a needlessly tedious task given that the branch analysis already contains this information. Moreover, as discussed in \cref{subsec:gate_charge}, the dispersive approximation fails even at low photon numbers for transitions between chargelike states near the top of the well, and this was shown to affect the ionization threshold. Thus, a perturbative approach cannot be expected to generally produce accurate results when the gate charge fluctuates.

Finally, let us stress that the order of the multiphoton process responsible for the swapping of the ground- or excited-state branch with any other branch can be diagnosed by considering the modular spectrum of the bare transmon states. For instance, here we knew that the four-photon process was responsible for the swapping of branches $B_{1_t}$ and $B_{7_t}$, since $(E_{7_t} - E_{1_t}- 4 \omega_r)/2\pi = \SI{1.98}{GHz} $ was the smallest detuning between these transmon states compared to any other multiphoton process. Thus, the order of these multiphoton resonances changes with the detuning $\Delta$. For instance, going from small to large negative detuning, the three peaks we see in the detuning sweep in \cref{fig:comparison_quantum_floquet}(a) are a consequence of multiphoton processes of fifth, fourth, and third order between the ground state and branches $B_{7_t}$, $B_{6_t}$, and $B_{5_t}$, respectively. 

\section{Effective drive on the transmon} \label{App. Sec.: Effective drive on the transmon}

In \cref{sec: Floquet with just transmon}, we obtain \cref{eq: driven Transmon Hamiltonian} by applying a displacement transformation $\hat D(\alpha) = \exp(\alpha \hat a^\dag-\alpha \hat a)$ on the driven qubit-resonator master equation [\cref{eq:Full_ME}]. In that transformation, the amplitude $\alpha$ is time dependent and takes the form $\alpha(t) = \int_0^t d\tau\,\mathcal E_d(\tau)e^{-(i\omega_r + \kappa/2)(t-\tau)}$. With that choice, the transformation eliminates the resonator drive from the Hamiltonian and induces an effective transmon drive $\mathcal E_t(t) \hat n_t = 2g\text{Im}[\alpha(t)]\hat n_t$. 

Taking the resonator drive to be $\mathcal E_d(t) = \varepsilon_d \sin(\omega_d t)$ with constant amplitude $\varepsilon_d$, we have that
\begin{equation}
\begin{split}
\alpha(t) =& \frac{\varepsilon_{d}}{2} \left[\frac{e^{-i\omega_{d}t} - e^{-(i\omega_r + \kappa/2)t}}{\omega_- -i\kappa/2} \right.\\
&\qquad - \left. \frac{e^{i\omega_{d}t} - e^{-(i\omega_r + \kappa/2)t}}{\omega_+ -i\kappa/2} \right],
\end{split}
\end{equation}
where $\omega_\pm = \omega_r \pm \omega_{d}$. Therefore, the amplitude of the transmon drive reads
\begin{equation} \label{eq: full transmon drive after displacement}
\begin{split}
\mathcal{E}_t(t) =& g \varepsilon_d \left[ \frac{\cos(\omega_d t - \phi_-) - \cos(\omega_r t - \phi_-)e^{-\kappa t/2}}{\sqrt{\omega_-^2 + (\kappa/2)^2}} \right. \\
&\quad -\left. \frac{\cos(\omega_d t + \phi_+) - \cos(\omega_r t - \phi_+)e^{-\kappa t/2}}{\sqrt{\omega_+^2 + (\kappa/2)^2}} \right],
\end{split}
\end{equation}
where $\tan \phi_\pm = -2\omega_\pm/\kappa$. Because $\omega_+ \gg g \varepsilon_d$, we can ignore the second line in the above expression. Furthermore, assuming that $|\omega_-|\ll \kappa$, the drive amplitude takes the simpler form
\begin{equation} \label{eq: simplified transmon drive}
\mathcal E_t(t) = \frac{2g \varepsilon_{d}}{\kappa}(1-e^{-\kappa t/2}) \cos(\omega_d t).
\end{equation}
This is the expression that we use in \cref{sec: Floquet with just transmon}. 

Beyond this limiting regime, the effective drive on the transmon [\cref{eq: full transmon drive after displacement}] can still be well approximated by a fast periodic drive with an envelope that grows slowly as $1-e^{-\kappa t/2}$, although not monotonically when $|\omega_r-\omega_d|$ is finite. In particular, for large $|\omega_r - \omega_d|$ compared to $\kappa$, an avoided crossing in the Floquet spectrum might be dynamically crossed multiple times.

An important remark is that, when $|\omega_r - \omega_d| \gtrsim \kappa$, we still obtain a better approximation to the full transmon-resonator dynamics by using \cref{eq: simplified transmon drive} than by using \cref{eq: full transmon drive after displacement} for the driven transmon. The reason is that the resonator responds at a qubit-state-dependent frequency that is typically closer to the readout frequency $\omega_d$ than to $\omega_r$. Thus, the monochromatic-tone approximation is more accurate.

\section{Floquet formalism}
\label{App. Sec.: Floquet formalism}
While for the numerical simulations we extract the Floquet Hamiltonian directly from the propagator over one period, $\hat U(T) \equiv e^{-i\hat H_F T}$, it is instructive to express the Floquet problem in Shirley space, also known as the infinite replicas picture~\cite{Shirley1965Solution}.

The infinite replicas picture is obtained by solving the time-dependent Schr\"odinger equation in frequency space. Defining the Floquet quasienergies and modes of the time-dependent periodic Floquet Hamiltonian via $\hat H_F \ket{\phi_{i_t}} = \epsilon_{i_t} \ket{\phi_{i_t}}$, a particular solution to the Schr\"odinger equation is the Floquet state $\ket{\psi_{i_t}(t)} = e^{-i\epsilon_{i_t}t}\ket{\phi_{i_t}(t)}$. The quasienergy is defined within the first Brillouin zone $\epsilon_{i_t} \in [-\omega_d/2, \omega_d/2]$. The Floquet modes are periodic in time, $\ket{\phi_{i_t}(t)} = \ket{\phi_{i_t}(t+T)}$, where $\ket{\phi_{i_t}(t)} = \hat{U}(t) e^{i \hat H_F t}\ket{\phi_{i_t}(0)}$ and $\ket{\phi_{i_t}} \equiv \ket{\phi_{i_t}(0)}$. 

In the case of the driven transmon, we may expand both the Hamiltonian $\hat H_t(t)$ in \cref{eq: driven Transmon Hamiltonian} and the Floquet modes in Fourier components as $\hat H_t(t) = \hat H_t + (\varepsilon_t/2) (e^{i\omega_dt} + e^{-i\omega_d t}) \hat{n}_t$ and $\ket{\phi_{i_t}(t)} = \sum_m e^{im\omega_d t}\ket{\phi_{i_t,m}}$, respectively. We can then substitute them into Schr{\"o}dinger's equation $i\partial_t \ket{\psi_t(t)} = \hat H(t)\ket{\psi_t(t)}$, leading to the eigenvalue problem
\begin{equation} \label{eq: Floquet replicas Hamiltonian}
\begin{split}
&(\hat H_t + m\omega_d)\ket{\phi_{i_t, m}} + \frac{\varepsilon_t}{2} \hat{n}_t (\ket{\phi_{i_t, m+1}} + \ket{\phi_{i_t, m-1}}) \\
&= \epsilon_{i_t}\ket{\phi_{i_t,m}}.
\end{split}
\end{equation}
The infinite replicas eigenvalue problem in \cref{eq: Floquet replicas Hamiltonian} can be interpreted more clearly if written in matrix form as
\begin{align} \label{eq: Shirley space equation}
%\begin{split}
&{\begin{pmatrix}
\ddots & \ddots & \ddots & \\
\ddots & \hat H_t - \omega_{d} & \frac{\varepsilon_t}{2}\hat n_t & 0 \\
\ddots & \frac{\varepsilon_t}{2} \hat n_t & \hat H_t & \frac{\varepsilon_t}{2}\hat n_t & \ddots \\
& 0 & \frac{\varepsilon_t}{2}\hat n_t & \hat H_t + \omega_{d} & \frac{\varepsilon_t}{2}\hat n_t \\
& & \ddots & \ddots & \ddots & \ddots 
\end{pmatrix} \begin{pmatrix}
\vdots \\ \ket{\phi_{i_t, -1}} \\ \ket{\phi_{i_t, 0}} \\ \ket{\phi_{i_t, 1}} \\ \vdots
\end{pmatrix}}\nonumber \\
&{\qquad = \epsilon_{i_t} \begin{pmatrix}
\vdots \\ \ket{\phi_{i_t, -1}} \\ \ket{\phi_{i_t, 0}} \\ \ket{\phi_{i_t, 1}} \\ \vdots
\end{pmatrix}.}
%\end{split}
\end{align}
Thus, the original time-dependent eigenvalue problem is mapped to a static eigenvalue problem involving infinite replicas of transmons ordered in a one-dimensional lattice. The transmon at site $m$ has a static Hamiltonian $\hat H_t + m\omega_{d}$ and is coupled to nearest-neighbor sites $m\pm 1$ via the ``tunneling'' operator $(\varepsilon_t/2) \hat n_t$. In the absence of a drive, $\varepsilon_t=0$, the Floquet mode's $m$th component is just $\ket{\phi_{i_t,m}[\varepsilon_t=0]} = \ket{i_t, m}$, i.e., the bare transmon state $i_t$ at site $m$. Moreover, in the replicas picture, the bare states $\ket{i_t, m}$ and $\ket{j_t, n}$ can hybridize at finite $\varepsilon_t$ provided that they can be connected via a tunneling path along the one-dimensional lattice. Moving back to the original time-dependent picture, the interpretation is that there can be a transition between levels $i_t$ and $j_t$ provided $|n-m|$ drive photons at frequency $\omega_{d}$ are absorbed ($n>m$) or emitted ($n<m$).

With a second-order perturbative expansion in $\varepsilon_t$, we can compute a correction to the bare ($\varepsilon_t=0$) frequencies $E_{i_t} + m \omega_d$. The ac-Stark shift of level $i_t$ is
\begin{align}
 \nonumber
 \chi_{i_t,m} =& \frac{\varepsilon_t^2}{4} \sum_{j_t} 
 \left(
 \frac{ |\bra{i_t}\hat n_t \ket{j_t}|^2}{E_{i_t, m} - E_{j_t, m+1}} + \frac{ |\bra{i_t}\hat n_t \ket{j_t}|^2}{E_{i_t, m} - E_{j_t, m-1}}
 \right)
 \\
 \nonumber =& \frac{\varepsilon_t^2}{2} \sum_{j_t} \frac{\omega_{i_tj_t} |\bra{i_t}\hat n_t \ket{j_t}|^2}{\omega_d^2 - \omega_{i_t j_t}^2} \\
 =& 2\bar{n}_r\sum_{j_t} \frac{\omega_{i_t j_t}|g_{i_t j_t}|^2}{\omega_d^2 - \omega_{i_t j_t}^2},
\end{align}
where we use the definitions $\omega_{i_t j_t} = E_{j_t} - E_{i_t}$, $g_{i_t j_t} = g \bra{i_t}\hat n_t \ket{j_t}$, and $\varepsilon_t = 2g \sqrt{\bar{n}_r}$. Note that this $\chi_{i_t, m}$ shift is independent of the replica index $m$, as expected, since all replicas are equivalent. Importantly, this expression is exactly the frequency shift of the transmon level $i_t$ obtained with the fully quantum model [\cref{eq:chi_shifts}] with an average number of photons $\bar n_r$ in the resonator and with $\omega_d = \omega_r$. This result is not accidental. Indeed, as we now explain, the Shirley Hamiltonian is equivalent in a rigorous way to a semiclassical version of the static Hamiltonian $\hat{H}_{tr}$. Furthermore, this result explains why the energies modulo $\omega_r$ of the branch eigenstates and the Floquet quasienergies are nearly identical; compare \cref{fig: fully quantum negative detuning}(b) with \cref{fig:Floquet negative detuning}(b) and \cref{fig:pos_branch_analysis}(b) with \cref{fig:Floquet positive detuning}(b).

To make the connection to \cref{eq: Floquet replicas Hamiltonian} more evident, we first apply a trivial phase shift to the creation and annihilation operators in $\hat{H}_{tr}$, which changes the capacitive coupling $-ig (\hat{n}_t-n_g)(\hat{a} - \hat{a}^\dagger) \to g (\hat{n}_t-n_g)(\hat{a} + \hat{a}^\dagger) $ but leaves the eigenenergies invariant. We then let
\begin{align}
|\phi_{i_t} \rangle \equiv \sum_{j_t,n_r} c^{(i_t)}_{j_t, n_r} \ket{j_t, n_r} \equiv \sum_{n_r}|\phi_{i_t,n_r}, n_r\rangle
\end{align}
denote a generic branch-$B_{i_t}$ eigenstate of $\hat{H}_{tr}$ with energy $\epsilon_{i_t}$. Here, $|\phi_{i_t,n_r}, n_r\rangle$ is a tensor product of the transmon state $|\phi_{i_t,n_r}\rangle \equiv \sum_{j_t} c_{j_t,n_r}^{(i_t)} |j_t\rangle$ with the $n_{r}$th Fock state. Given that $|\phi_{i_t}\rangle$ is an eigenvector of $\hat{H}_{tr}$, we can conclude that the various $|\phi_{i_t,n_r}\rangle$ must satisfy
\begin{align} \nonumber
&(
\hat{H}_t + n_r \omega_r
)
|\phi_{i_t, n_r}\rangle
+
\frac{\varepsilon_{t, n_r+1}}{2}(\hat{n}_t-n_g)
|\phi_{i_t, n_r+1}\rangle \\
&+
\frac{\varepsilon_{t, n_r}}{2}(\hat{n}_t-n_g)
|\phi_{i_t, n_r-1}\rangle
=
\epsilon_{i_t} |\phi_{i_t, n_r}\rangle,
\label{eq:quantum_Shirley_Floquet_parallel}
\end{align}
where we introduce the function $\varepsilon_{t, n_r} \equiv 2 g \sqrt{n_r}$.

To render the fully quantum eigenvalue problem in \cref{eq:quantum_Shirley_Floquet_parallel} equivalent to \cref{eq: Floquet replicas Hamiltonian}, we assume that the eigenvectors of $\hat{H}_{tr}$ have a large average photon number $\bar{n}_r$ and that they are localized in Fock space. This allows us to ignore the photon-number-dependent variation in the matrix elements $\varepsilon_{t, n_r} \approx \varepsilon_{t,n_r+1} \approx \varepsilon_{t, \bar{n}_r}$. This also allows us to ignore the existence of the vacuum state. Indeed, in the semiclassical picture, one can absorb or emit an arbitrary number of photons, which is what leads to the replica index $m$ spanning all integers. In contrast, in the quantum picture, the transmon cannot absorb an arbitrary number of photons. With the above assumption, however, the existence of the vacuum is irrelevant, and we can safely ignore the lower bound on $n_r$. Note that these approximations become better at large photon numbers or, equivalently, at strong drives. In that limit, therefore, the two eigenvalue problems in \cref{eq:quantum_Shirley_Floquet_parallel} and \cref{eq: Floquet replicas Hamiltonian} are equivalent up to the dependence on $n_g$. It can then be shown that the dependence on $n_g$ can be removed with the gauge transformation $|\tilde{\phi}_{i_t,k_r}\rangle = \sum_{n_r} J_{n_r-k_r}\left(\frac{\varepsilon_t n_g}{\omega_d}\right) |\phi_{i_t, n_r}\rangle$, where $J_k(x)$ is the Bessel function of the first kind, corresponding to the gauge transformation that removes an irrelevant constant $-\varepsilon_t n_g \cos(\omega_d t)$ from the transmon drive.

\section{Floquet-Landau-Zener transition probability}
\label{App. subsec: Floquet-Landau-Zener transition probability}

The Landau-Zener formula for the transition probability at an avoided crossing between two Floquet quasienergies gives accurate predictions, even for fast pulses~\cite{Drese1999Floquet, Ikeda2022Floquet-Landau-Zener}. In our mapping of the driven transmon-resonator Hamiltonian to a Floquet problem, we do not directly change the detuning of the quasienergies linearly in time, as would be the case for the standard Landau-Zener argument. Instead, the relevant parameters are the amplitude $\varepsilon_d$ of the drive on the resonator and the resonator decay rate $\kappa$, both of which change the effective drive amplitude $\varepsilon_t$ on the transmon following \cref{eq: simplified transmon drive}. That effective drive amplitude is the quantity which controls the detuning between the quasienergies.

To deduce the effective speed $v$ at which resonances are traversed as a function of the drive amplitude $\varepsilon_t(t)$, we follow Ref.~\cite{Ikeda2022Floquet-Landau-Zener} in approximating the difference in the quasienergies close to an avoided crossing as
\begin{equation}\label{eq:quasi-energy-detuning}
\begin{split}
\epsilon_{j_t}[\varepsilon_t(t)] - \epsilon_{i_t}[\varepsilon_t(t)] &\approx \sqrt{\Delta_{\rm ac}^2 + [v(t-t_{\rm ac})]^2} \\
&\approx \Delta_{\rm ac} + \frac{1}{2} \frac{[v(t-t_{\rm ac})]^2}{\Delta_{\rm ac}}.
\end{split}
\end{equation}
In this expression, $t_{\rm ac}$ is the time at which the minimum gap $\Delta_{\rm ac}$ in the avoided crossing is reached. The value of the drive amplitude at that time is denoted $\varepsilon_{t, {\rm ac}} \equiv \varepsilon_t(t_{\rm ac})$. Expanding the left-hand side of \cref{eq:quasi-energy-detuning} to second order (the first derivatives vanish at the avoided crossing), we obtain
\begin{equation}
\begin{split}
&\epsilon_{j_t}[\varepsilon_t(t)] - \epsilon_{i_t}[\varepsilon_t(t)] \approx \epsilon_{j_t}[\varepsilon_{t,{\rm ac}}] - \epsilon_{i_t}[\varepsilon_{t, {\rm ac}}] \\ 
&+ \bigg| \frac{d^2 \epsilon_{j_t}[\varepsilon_t]}{d \varepsilon_t^2}\bigg|_{\varepsilon_{t, {\rm ac}}} \left(\frac{d\varepsilon_t(t)}{dt}\bigg|_{t_{\rm ac}}\right)^2 (t-t_{\rm ac})^2.
\end{split}
\end{equation}
Combining the two above expressions, we find that the speed $v$ is given by the relation
\begin{equation}
\frac{v^2}{2\Delta_{\rm ac}} = \bigg| \frac{d^2 \epsilon_{j_t}[\varepsilon_t]}{d \varepsilon_t^2}\bigg|_{\varepsilon_{t, {\rm ac}}} \left(\frac{d\varepsilon_t(t)}{dt}\bigg|_{t_{\rm ac}}\right)^2.
\end{equation}
Note that, here, we assume that the second derivatives of $\epsilon_{i_t}[\varepsilon_t]$ and $\epsilon_{j_t}[\varepsilon_t]$ are equal at the avoided crossing.

Knowing the gap size $\Delta_{\rm ac}$ and the speed $v$, the Landau-Zener diabatic transition probability can be computed as $P_{\rm LZ} = \exp\left(-\pi \Delta_{\rm ac}^2/2v\right)$. For the parameters used in \cref{fig:Floquet negative detuning}(c), we obtain $P_{\rm LZ}=0.35$ for the diabatic crossing and $1-P_{\rm LZ} = 0.65$ for the adiabatic crossing. The quasienergy gap and the speed are numerically obtained from the tracked quasienergies and from the functional form of the drive amplitude $\varepsilon_t(t)$.

\section{Comparison of the Bohr-Sommerfeld energies with Floquet theory \label{App. Sec.: Bohr-Sommerfeld vs Floquet}}

\begin{figure}[t!]
 \centering
 \includegraphics[width=\linewidth]{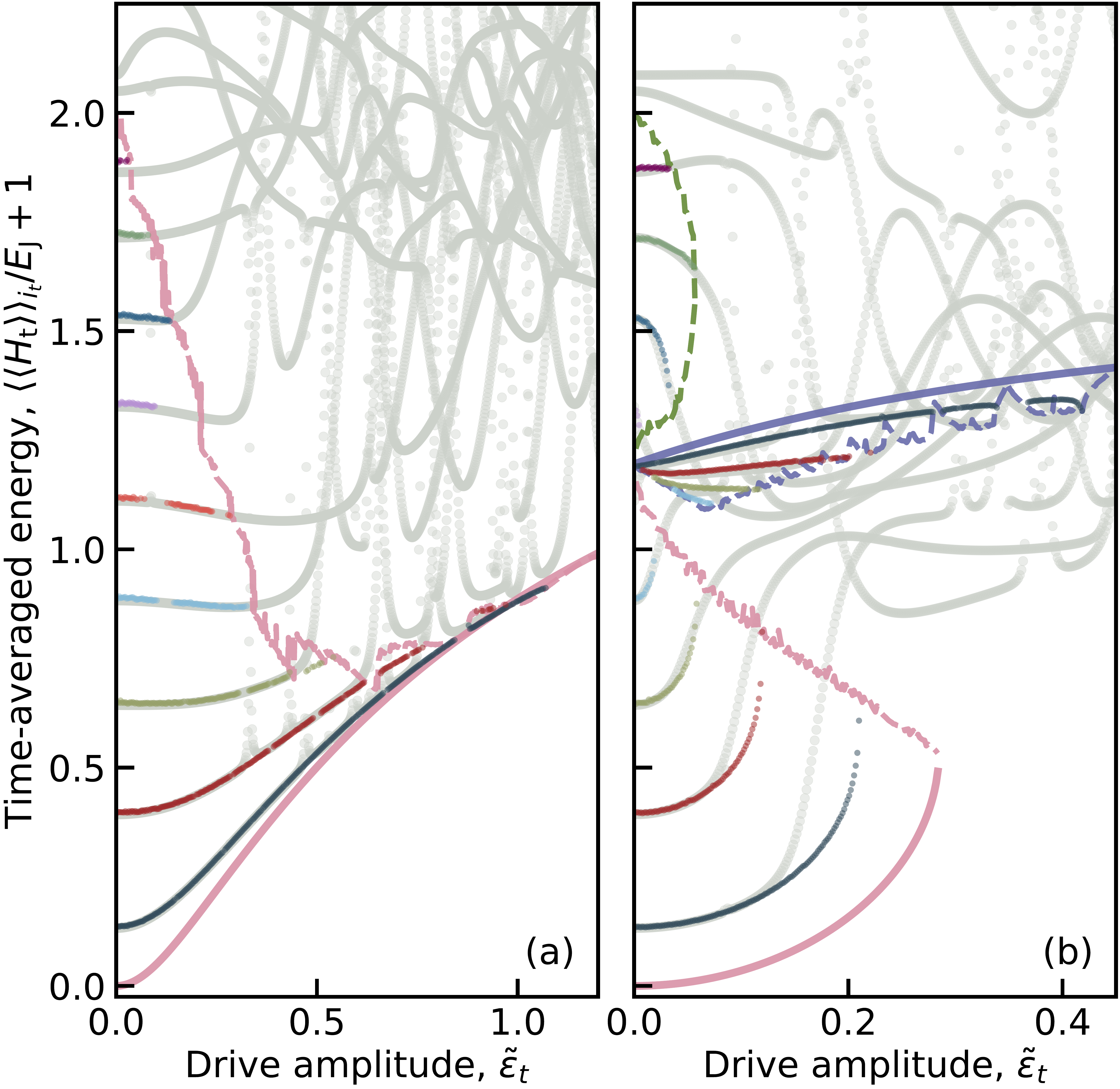}
 \caption{
 Time-averaged energies $\langle\langle H_t \rangle\rangle_{i_t}$ of the Bohr-Sommerfeld orbits (colored dots with the same color code as in \cref{fig:BohrSommerfeld_drive_heatmaps}) and time-averaged energies $\langle\langle \hat{H}_t \rangle \rangle_{i_t}$ of the Floquet modes at $n_g = 0$ (gray dots) as a function of the rescaled drive amplitude $\Tilde{\varepsilon}_t$. The solid lines indicate the energies of the center of the main regular region (pink) and of the $\nm{1}{1}$ resonance (indigo). The dashed lines indicate the energy of the boundaries of the regular regions, namely, the main regular region (pink), the $\nm{1}{1}$ resonance (indigo), and the innermost and outermost orbits external to the $\nm{1}{1}$ resonance at positive detuning (green). We show (a) the case of negative detuning ($\Tilde{\omega}_d = 1.152$) and (b) the case of positive detuning ($\Tilde{\omega}_d = 0.807$). The Bohr-Sommerfeld $\langle\langle H_t \rangle\rangle_{i_t}$ are obtained by averaging the Hamiltonian over $\mathcal{N}=1000$ periods using $40$ points per period. The Floquet $\langle\langle \hat{H}_t \rangle \rangle_{i_t}$ are obtained using $31$ points per period. The drive amplitude is varied in increments of $\delta\varepsilon_t/2\pi = \omega_p \delta\Tilde{\varepsilon}_t/2\pi = \SI{10}{MHz}$. The Bohr-Sommerfeld orbits and the region boundaries are identified using the procedure described in \cref{App. Sec.: Area algorithm chaos}.
 }
 \label{fig:FloquetBohrSommerfeld_energies}
\end{figure}

In this appendix, we show that the Bohr-Sommerfeld quantization procedure gives a sensible approximation to quantum theory for typical transmon parameters. This is done by comparing the results obtained from the Bohr-Sommerfeld quantization approach to the Floquet analysis developed in \cref{sec: Floquet with just transmon} for $n_g = 0$. More precisely, we compare the time-averaged transmon energy over the coordinates $\{\varphi_{i_t}(\tau), n_{i_t}(\tau)\}$ for the Bohr-Sommerfeld orbits:
\begin{equation} \label{eq:bs_time_average} 
 \langle\langle H_t \rangle\rangle_{i_t} = \lim_{t \rightarrow \infty} \frac{1}{t} \int_{0}^t d\tau \, H_t\bigl(\varphi_{i_t}(\tau), n_{i_t}(\tau)\bigr),
\end{equation}
to the time-averaged energy of the Floquet modes~\cite{Cohen2023Reminiscence}:
\begin{equation} \label{eq:floquet_time_average}
 \langle\langle \hat{H}_t \rangle \rangle_{i_t} = \frac{1}{T}\int_{0}^{T} d\tau \, \bra{\phi_{i_t}(\tau)} \hat{H}_t \ket{\phi_{i_t}(\tau)},
\end{equation}
as shown in \cref{fig:FloquetBohrSommerfeld_energies}(a,b). In these plots, the gray dots correspond to the Floquet mode energies given by \cref{eq:floquet_time_average} and the colored dots to the time-averaged energies of the Bohr-Sommerfeld orbits given by \cref{eq:bs_time_average}. For the latter, we use the same color code as in \cref{fig:BohrSommerfeld_drive_heatmaps}. The scale of the vertical axis is chosen such that zero energy corresponds to the bottom of the cosine well while $2E_J$ corresponds to the energy of the top of the well, i.e., the energy of the separatrix of $H_t$. The significance of the colored solid and dashed lines is explained below. 

Focusing first on the case of negative detuning [\cref{fig:FloquetBohrSommerfeld_energies}(a)], we observe that the average energy of the Floquet modes (gray dots) located near the bottom of the well changes smoothly with drive amplitude at small drive amplitudes. By contrast, modes located near the separatrix rapidly hybridize, resulting in erratic behavior with the drive amplitude. In Ref.~\cite{Cohen2023Reminiscence}, this was shown to be a signature of chaotic behavior. As the drive amplitude is further increased, the chaotic layer widens until it absorbs the Floquet modes associated with the qubit logical states. Note that the resonances responsible for ionization can be seen as sharp features of the average energy connecting Floquet modes at the bottom of the well with modes closer to the top of the well. 

Before chaotic behavior emerges, the average energies of the Bohr-Sommerfeld orbits (colored dots) match the average Floquet energies very well. It is interesting to note that this is the case even though there are only a few levels in the well. Thus, the classical method is quite robust. Moreover, the discontinuities that are observed in the average energies of the Bohr-Sommerfeld orbits coincide with the ranges of $\tilde{\varepsilon}_t$ for which the corresponding orbits cease to exist because of collisions with $\nm{n}{m}$ resonances. The solid and dashed pink lines denote the average energy of the orbit at the center of the main regular region [corresponding to the gray line in \cref{fig:BohrSommerfeld_drive_heatmaps}(a)] and the average energy of the outermost orbit of the main regular region, respectively. These two lines delimit the energy window into which regular orbits exist, i.e., where Bohr-Sommerfeld states are supported. The correspondence between the dashed pink line and the region where the average Floquet energy shows chaotic behavior is striking.

At positive detuning, we again see an excellent agreement between the classical and the Floquet predictions at small drive amplitudes; see \cref{fig:FloquetBohrSommerfeld_energies}(b). Notably, both approaches show the emergence of a set of states in the vicinity of $i_t = 4$ whose energies change smoothly with drive amplitude. This is the bunching layer discussed in \cref{subsec: fully quantum positive detuning} and \cref{subsec: floquet positive detuning} and which is caused by a large $\nm{1}{1}$ resonance as discussed in \cref{subsec:classical_positive_det}. Chaotic behavior emerges not only close to the separatrix of $\tilde H_t$, but also close to the bunching layer. This is the behavior expected from \cref{fig:poincare_maps_2x4,fig:BohrSommerfeld_drive_heatmaps}, and it results from the appearance of a new separatrix close to the $\nm{1}{1}$ resonance. Increasing further the drive amplitude, the chaotic layer widens until all states inside the cosine potential become chaotic. 

Because of the presence of the large $\nm{1}{1}$ resonance, there are three different regions that can support Bohr-Sommerfeld states, namely, the main regular region, the $\nm{1}{1}$ resonance itself, and the orbits surrounding both the main regular region and the $\nm{1}{1}$ resonance; see \cref{fig:BohrSommerfeld_drive_heatmaps}(d). In \cref{fig:FloquetBohrSommerfeld_energies}(b), the solid and dashed pink lines correspond to the average energies of the orbits at the center and inner edge of the main regular region, respectively. Furthermore, the solid and dashed indigo lines are the average energies of the center and inner edge of the $\nm{1}{1}$ resonance, respectively. The lower and upper dashed green lines are the average energies of the innermost and outermost orbits external to the $\nm{1}{1}$ resonance, respectively. The agreement between the Bohr-Sommerfeld energies and Floquet modes energies in the three identified zones is excellent. We verified that the correspondence between the Floquet states and the regular orbits holds better as $\hbar_{\rm{eff}}$ is made smaller (not shown). In the limit $\hbar_{\rm{eff}} \rightarrow 0$ ($E_J/E_C \rightarrow \infty$), the lowest-energy Bohr-Sommerfeld state coincides with the energy of the center of the main regular region and ionizes at around $\Tilde{\varepsilon}_t \sim 0.3$, where the solid and dashed pink lines meet in \cref{fig:FloquetBohrSommerfeld_energies}(b). Consequently, the Bohr-Sommerfeld critical photon number associated with the $\nm{1}{1}$ resonance has a saturating value.

Finally, we note that the center of the $\nm{1}{1}$ resonance is a point of stability with higher average energy than the surrounding points. Thus, the bunched states are excitations within an effective inverted potential well generated by the drive. Such an inverted effective potential well bears a superficial resemblance to the Kapitza pendulum, i.e., a rigid pendulum stabilized ``upside-down'' by periodically changing the height of the pivot point. With the $\nm{1}{1}$ resonance of interest here, however, the stabilized orbits oscillate around $\varphi=0$ and not around $\varphi=\pm \pi$. 

\section{Calculation of the Bohr-Sommerfeld critical photon numbers \label{App. Sec.: Area algorithm chaos}}

In this appendix, we describe the method used to compute the Bohr-Sommerfeld critical photon numbers $\{n_{\mathrm{crit},i_t}^{c}\}$ introduced in \cref{subsec:bohr_sommerfeld_critical_photon_numbers}. The key is to determine whether the Bohr-Sommerfeld orbit associated with a given qubit state $i_t$ exists. If the orbit is chaotic or overlaps with a $\nm{n}{m}$ resonance, there exists no closed and fully connected regular orbit with the required area, and the qubit state is then declared unstable. The various photon numbers at which the Bohr-Sommerfeld orbit associated with qubit state $i_t$ becomes nonexistent yield the Bohr-Sommerfeld critical photon numbers $\{n_{\mathrm{crit},i_t}^{c}\}$. 

For a given value of the rescaled drive frequency $\tilde{\omega}_d$ (i.e., a fixed value of the detuning $\Delta$) and of the rescaled drive amplitude $\tilde{\varepsilon}_t$, we solve Hamilton's equations to obtain the orbits over $\mathcal{N}$ drive periods for all initial conditions along the axis $(\Tilde{\varphi}_t = 0, \Tilde{n}_t^0)$, with $\Tilde{n}_t^0 \in [-2.0, 2.0]$ varied in increments of $1/300$. We then attempt to compute the areas of all orbits using Green's theorem. More precisely, we choose some phase-space field $\mathbf{F} = \left\{ P(\tilde{\varphi}_t, \tilde{n}_t), Q(\tilde{\varphi}_t, \tilde{n}_t) \right\}$ such that $\nabla \times \mathbf{F}$ is a unit vector perpendicular to the phase-space plane. With that choice, Green's theorem yields
 \begin{equation}
 \begin{split}
  \mathcal{A} &= \iint_\mathcal{D} (\nabla \times \mathbf{F})\cdot d\mathbf{A} = \oint_\mathcal{C}\mathbf{F}\cdot \mathbf{dr} \\
  &\approx \sum_\nu \left[ P(\tilde{\varphi}_t^\nu, \tilde{n}_t^\nu) \Delta \Tilde{\varphi}_t^\nu + Q(\tilde{\varphi}_t^\nu, \tilde{n}_t^\nu) \Delta\Tilde{n}_t^\nu \right],
  \label{eq:Area_Green_general}
 \end{split}
 \end{equation}
 where the sum is over all points $\nu$ in the orbit and where $\Delta \Tilde{\varphi}_t^\nu = \Tilde{\varphi}_t^\nu - \Tilde{\varphi}_t^{\nu-1}$ and $\Delta \Tilde{n}_t^\nu = \Tilde{n}_t^\nu - \Tilde{n}_t^{\nu-1}$ are the coordinate differences between adjacent points of the orbit. A simple choice for $\mathbf{F}$ is $P(\tilde{\varphi}_t,\tilde{n}_t) = 0$ and $Q(\tilde{\varphi}_t,\tilde{n}_t) = \tilde{\varphi}_t$, yielding
 \begin{equation}
  \mathcal{A} \approx \left| \sum_\nu \Tilde{\varphi}_t^\nu \Delta \Tilde{n}_t^\nu \right|.
  \label{eq:Area_Green_discrete}
 \end{equation}
The absolute value is present to obtain a positive area independently of the integration direction along the orbit. In general, the points of an orbit in the Poincaré section are not naturally ordered along the orbit, because they are obtained stroboscopically. Therefore, we must first order the points of an orbit to compute an area using \cref{eq:Area_Green_discrete}. If most points of an orbit can be ordered on a closed, connected, and near-continuous orbit, an area can be successfully calculated and a closed and connected orbit with that area can exist. In all other cases, such an orbit cannot exist.
 
In the attempt to order the points along an orbit, we start at the initial point of the orbit. We then find the point of the orbit that is closest to the initial point in the Poincaré section. The nearest neighbor of this second point is then selected, excluding the points that were already selected. The procedure is repeated until it returns to the initial point or until the nearest neighbor is farther than a distance threshold $d$ chosen as
 \begin{equation}
  d =
  \begin{cases}
  \frac{2\pi r_c m}{\mathcal{N}} & \text{if}\quad\frac{\mathcal{N}d_{\rm{min}}}{2\pi m} < r_c < \frac{\mathcal{N}d_{\rm{max}}}{2\pi m}, \\
  d_{\rm{min}} & \text{if} \quad r_c \leq \frac{\mathcal{N}d_{\rm{min}}}{2\pi m}, \\ 
  d_{\rm{max}} & \text{if} \quad r_c \geq \frac{\mathcal{N}d_{\rm{max}}}{2\pi m} \text{ or no main region.} 
  \end{cases}
 \label{eq:sorting_threshold}
 \end{equation}
Here, $m$ is a scalar and $r_c$ is the distance between the initial condition $(0, \Tilde{n}_t^0)$ and the center of the main regular region (determined as described in the next paragraph). The first condition in \cref{eq:sorting_threshold} sets the distance threshold in proportion to the expected average distance between orbit points. Indeed, the quantity $2\pi r_c/\mathcal{N}$ is the arc length between orbit points distributed uniformly on a circle of radius $r_c$. Thus, $m$ is the number of average arc lengths tolerated between nearest neighbors. The second and third conditions in \cref{eq:sorting_threshold} set a minimum and maximum distance threshold, respectively. The minimum distance threshold $d_{\rm{min}}$ ensures that the threshold $d$ remains finite very close to the center of the main regular region, although such small orbits are not relevant for the parameter regime explored in this work (i.e., the area of a circle of radius $\mathcal{N}d_{\mathrm{min}}/2\pi m$ is smaller than the minimum area $\pi \hbar_{\rm eff}$ of a Bohr-Sommerfeld orbit). The maximum distance threshold $d_{\rm{max}}$ ensures that the threshold remains finite far from the main regular region or when the main regular region stops existing. This mitigates the number of chaotic orbits for which the algorithm occasionally finds a closed connected orbit [these appear as ``noise'' in the white regions of \cref{fig:BohrSommerfeld_drive_heatmaps}(a,b)]. The algorithm records a successful area calculation if it returns to the initial point and if the number of ordered points is larger than $\mathcal{N}/2$, and it otherwise records a failure. The latter condition prevents mistakenly identifying a single disconnected island of a $\nm{n}{m}$ resonance as a closed connected orbit. For the above procedure, we use $\mathcal{N} = 1000$, $2\pi m/\mathcal{N} \approx 0.65$, $d_{\rm{min}} = 0.10$, and $d_{\rm{max}} = 0.80$, which give us fast and reliable results.
 
Setting the threshold $d$ using \cref{eq:sorting_threshold} requires knowledge of the position of the center of the main regular region at all drive amplitudes. At $\Tilde{\varepsilon}_t=0$, the center sits at $(0,0)$ in phase space. If the drive amplitude is slightly increased, the center slightly moves away from its former position, such that an orbit with initial condition $(0,0)$ now encloses the new center. The position of the new center is estimated as the center of mass of that orbit. The method is iterated until the main regular region stops existing. This is detected by the orbit wandering too far away from the previous center position. To track the center, we increment $\Tilde{\varepsilon}_t$ by $0.01$, use orbits with $\mathcal{N}=1500$ periods, and set the wandering-off threshold distance to $1.0$. In \cref{fig:BohrSommerfeld_drive_heatmaps}, the center of the main regular region is shown as gray lines starting at $(0, 0)$ at zero drive amplitude in \cref{fig:BohrSommerfeld_drive_heatmaps}(a,b), while it appears as gray dots in \cref{fig:BohrSommerfeld_drive_heatmaps}(c,d).

The above procedure allows us to compute the phase-space area of orbits as a function of $\Tilde{\omega}_d$ and $\Tilde{\varepsilon}_t$. We first vary $\Tilde{\omega}_d$ [$\Delta/2\pi$] between $0.5$ [$\SI{3.02}{GHz}$] and $1.5$ [$\SI{-3.49}{GHz}$] in increments of $0.01$ but exclude very small detunings $\Tilde{\omega}_d \in (0.90,1.03)$ [$\Delta/2\pi \in (-0.47, 0.41)\,\textrm{GHz}$] where the dispersive approximation does not hold. For each value of $\Tilde{\omega}_d$, we then vary $\tilde{\varepsilon}_t$ in increments of $\delta\tilde{\varepsilon}_t = \delta \varepsilon_t/\omega_p = 0.0015323$, corresponding to $\delta \varepsilon_t/2\pi = \SI{10}{MHz}$. The results are shown in \cref{fig:BohrSommerfeld_drive_heatmaps} for (a) negative detuning and (b) positive detuning. The colored regions indicate the initial conditions where an area was successfully calculated. In particular, the Bohr-Sommerfeld orbits are highlighted as pale colored lines. The dark-colored regions represent the extent of quantum fluctuations around a given Bohr-Sommerfeld orbit. More precisely, they contain all orbits with an area within $\pi \hbar_{\rm eff}$ of that orbit at fixed $\Tilde{\varepsilon}_t$. The white regions indicate the initial conditions where the algorithm failed to calculate an area. The Bohr-Sommerfeld critical photon numbers $\{n_{\mathrm{crit},i_t}^{c}\}$ are the points where the Bohr-Sommerfeld line for state $i_t$ collides with the white regions as the drive amplitude is increased. These points are represented by stars in \cref{fig:BohrSommerfeld_drive_heatmaps}. A lower bound to the $\{n_{\mathrm{crit},i_t}^{c}\}$ from quantum fluctuations is obtained from the points where the boundaries of the dark-colored regions collide with the white regions as the drive amplitude is increased. These points are not shown in \cref{fig:BohrSommerfeld_drive_heatmaps}(a,b).

Finally, we quantify the strength of the resonances at the critical photon numbers by the area they occupy in phase space; see \cref{fig:Floquet_Nt_CAncrits}(a) in the main text. To obtain the area of a resonance, we start at a critical point in \cref{fig:BohrSommerfeld_drive_heatmaps} (star) and vary $\Tilde{n}_t^0$ across the resonance at fixed $\Tilde{\varepsilon}_t$. The area of the resonance is then obtained as the difference of the areas of the two orbits that jointly enclose the resonance. We repeat this procedure for every relevant resonance. Note that for degenerate pairs of resonances, such as the two $\nm{3}{2}$ resonances shown in \cref{fig:poincare_maps_2x4}(b), the procedure computes the total area of the pair.

%%%%%%%%%%%%%%%%%%%%%%%%%%%%%%%%
%%%%%%%% References %%%%%%%%%%
%%%%%%%%%%%%%%%%%%%%%%%%%%%%%%%%

\bibliography{Solving_Readout}

\end{document}